\documentstyle[12pt]{article}
\newcommand{\h}{\frac{\dot{H}}{H}}
\newcommand{\g}{\frac{G'}{G}}
\newcommand{\f}{\frac{F'}{F}}
\newcommand{\p}{\frac{P'}{P}}
\newcommand{\w}{W'}

\newcommand{\Qf}{\frac{{F'}^2}{F^2}}
\newcommand{\Qp}{\frac{{P'}^2}{P^2}}
\newcommand{\Dg}{\frac{G''}{G}}
\newcommand{\Df}{\frac{F''}{F}}
\newcommand{\Dp}{\frac{P''}{P}}
  
\newtheorem{proposition}{Proposition}
\newtheorem{lemma}{Lemma}
\begin{document}
\title{$G_2$ Perfect-Fluid Cosmologies with a proper conformal Killing
vector}
\author{ 
Marc Mars\thanks{Also at Laboratori de F\'{\i}sica Matem\`atica,
IEC, Barcelona.} \hspace{0.8mm} and Thomas Wolf \\ 
School of Mathematical Sciences, \\
Queen Mary and Westfield College, \\
University of London, \\
Mile End Rd, London, E1 4NS, U.K,}
\maketitle
\begin{abstract}
We study the Einstein field equations for
spacetimes admitting a maximal two-dimensional abelian group of
isometries acting orthogonally transitively on spacelike surfaces and,
in addition, with at least one conformal Killing vector. 
The three-dimensional conformal group is restricted to the case when the
two-dimensional abelian isometry subalgebra is an ideal
and it is also assumed to act on non-null hypersurfaces (both,
spacelike and timelike cases are studied).
We consider both, diagonal and non-diagonal metrics
and find all the perfect-fluid solutions under these
assumptions (except those already known). We find four families of
solutions, each one containing arbitrary parameters for which no
differential equations remain to be integrated. We write the
line-elements in a simplified form  and perform a detailed
study for each of these solutions, giving the kinematical quantities of the
fluid velocity vector, the energy-density and pressure, values of
the parameters for which
the energy conditions are fulfilled everywhere, the Petrov
type, the singularities in the spacetimes and the
Friedmann-Lema\^{\i}tre-Robertson-Walker metrics contained in each
family. 
\end{abstract}
\newpage

\section{Introduction}
One of the most successful ways of finding exact solutions of Einstein field 
equations has been the assumption of a certain degree of symmetry 
in the spacetime manifold \cite{KMSH}.
Although a few solutions are known
without any symmetry at all, a systematic study of exact
solutions with a given isometry group has only been performed 
when the dimension $r$ of the isometry group $G_r$ is $r\geq 2$. In the
particular case of $G_2$ spacetimes, it is usually
assumed that the isometry group is abelian
and acts orthogonally
transitively on 2-surfaces (that is to say, that the two planes
orthogonal to the group orbits at each point are themselves surface-forming).
In the cosmological context, the two-dimensional orbits of the isometry
group are taken spacelike ($S_2$) and the matter content is usually
assumed to be a perfect fluid.
The Einstein field equations in this situation are still very
complicated and further assumptions are needed to handle the problem.
They may be very different in nature, for instance:
separability of the
metric coefficients, degenerate Petrov types for the Weyl tensor, 
particular equations of state, kinematical properties of the
fluid velocity vector, Kerr-Schild ansatz, etc.  
Among these additional simplifications, we will assume the existence
of conformal symmetries
in the spacetime (for a definition of conformal symmetry see e.g. 
\cite{KMSH} and for a detailed study of some general properties of 
conformal Killing vectors see \cite{Hall}).

A conformal motion
in the manifold is interesting for several reasons. First of all,
it is a well-posed
geometrical assumption which, therefore, does not depend on the 
coordinate system we are using. The Friedmann-Lema\^{\i}tre-Robertson-Walker
(FLRW) spacetimes (which are at present our best candidates
to describe the real
universe, despite them being highly symmetric) possess a fifteen-dimensional
conformal group. Consequently, the inhomogeneous and anisotropic models
containing a conformal Killing vector
(CKV) are likely to contain FLRW limits
and, therefore, are suitable as generalizations of FLRW cosmologies
(in particular, they can be used 
to test several perturbative schemes used to study
inhomogeneities in the universe). 
Another reason for studying inhomogeneous cosmologies
admitting a CKV is the Ehlers-Geren-Sachs result \cite{EGS} which
implies that observers measuring an isotropic microwave background
radiation can exist only if the spacetime possesses a timelike
CKV and the observer moves collinearly with it.

Perfect-fluid spacetimes admitting a CKV 
have received considerable attention in the last few years. The case in which
the CKV is inheriting (meaning that the conformal
motion maps the fluid world-lines into themselves) has been studied in
\cite{CoTu1}. In particular, the spherically symmetric (\cite{CoTu2}
and references therein) and the
plane symmetric spacetimes \cite{CoCza} admitting an inheriting CKV
have been fully investigated. In \cite{Co} it was shown that
perfect-fluid spacetimes satisfying an equation of state $p=p(\rho)$
and admitting a CKV parallel to the 
fluid velocity must be FLRW models. 
A first attempt to perform
a systematic classification of spacetimes using its conformal
isometry group has been performed
in {\cite{CaAm} where some new exact solutions
have also been found. All this work showed that the perfect-fluid models
admitting a CKV were rare but not impossible. 
In vacuum metrics, the most general spacetime admitting a CKV \cite{EIMM} 
is either Minkowski or some type N solutions (representing
certain plane wave solutions). 

The perfect-fluid 
case when the spacetime admits an abelian two-dimensional isometry group
acting orthogonally transitively and one linearly independent
proper conformal motion 
(the homothetic
case was studied in \cite{Ali2}) 
has been considered only very recently \cite{AlTs}, \cite{COA}. In these
articles, the authors 
classify the three-dimensional conformal group according to the
possible inequivalent Lie algebras and find canonical
line-elements for each case.
Then, they concentrate on the diagonal metrics (which already excludes
a number of Lie algebras, see below) and study systematically the
possible solutions of Einstein's field equations for some selected
Lie algebras (in our notation below, expression (\ref{LieA}),
the Lie algebras considered
in {\cite{COA} correspond to Lie Algebra A with $b=0$).  

In this paper we will complete this work 
by considering in full generality the case of spacetimes admitting an
abelian orthogonally transitive $G_2$ with spacelike orbits
and one proper
conformal Killing vector\footnote{Obviously, the conformal Killing vector is
not unique since it is defined up to linear combinations of the Killing
vectors and a non-vanishing constant factor. 
Throughout the paper we will refer to this equivalence class as
'the conformal Killing vector'.} such that the
three-dimensional conformal
group has either timelike or spacelike orbits (for null orbits see
\cite{AlTs}). The final restriction we make is considering only
the three-dimensional conformal Lie algebras for which the abelian
isometry subalgebra is an ideal
(i.e.\ the
Lie bracket of the CKV with any of the two Killing vectors
is a linear combination of only the two Killing vectors).
We will not assume that the metric is diagonal 
(thus, we will study both classes
B(i) and B(ii) in Wainwright's
classification \cite{Wa}). Obviously, we will assume $b\neq 0$ in 
expression (\ref{LieA}) below when
restricting ourselves to the diagonal metrics. Although the assumption
of two Killing vectors (KVs) and one proper CKV
is the same as in \cite{MS1}, \cite{MS2}, in which
the stationary and axisymmetric case was analyzed, the
structure of the Lie algebras and differential equations to be solved turn
out to be rather different.

The plan of the paper is as follows. In section 2 we write down the
Einstein field equations in non-comoving coordinates
when the spacetime admits an abelian $G_2$ acting
orthogonally transitively on spacelike orbits.
Then, we establish the necessary and sufficient conditions for a 
solution of the system of Einstein equations to represent a perfect fluid.
We finish the section by showing that a new metric obtained by interchanging
the coordinates $t$ and $x$ in all the metric functions of a given solution
of the Einstein field equations has still the  energy-momentum
tensor with the structure of perfect fluid but with the 4-velocity
switching from timelike to tachyonic and vice versa.
This result allows us
to simplify the work considerably as only one half of the cases need
to be considered. In section 3 we write down the inequivalent
three-dimensional conformal Lie algebras for which the isometry abelian
subalgebra is an ideal. We then give the canonical
line-elements for each inequivalent case
(some of the forms of the Lie algebras and the line-elements are
different to those in \cite{COA} although they are completely equivalent). In
section 4 we study the non-diagonal metrics and proof that {\it no}
perfect-fluid solutions with this specification are possible.
Thus, in section 5 we
concentrate on the diagonal metrics (when $b\neq 0$, see above) and find
the general solution of Einstein's field equations. There exist four
families of solutions (each of them containing arbitrary parameters). These
families are written explicitly in a simplified form in section 6. Readers
interested mainly in the resulting metrics may advance to this section where
we give the Petrov type of the spacetimes,
the form of the fluid velocity vector, the shear tensor components, the
acceleration and the expansion of the fluid (vanishing rotation is
already a consequence of imposing an orthogonally transitive $G_2$).
The expressions for the
energy-density
and pressure of the fluid are written down and we find for which values
of the parameters the energy conditions are fulfilled. We also 
discuss the existence of an
equation of state $p=p(\rho)$,
the singularity structure of the spacetime and the
FLRW limit cases for each of the families.

\section{Some results on general abelian $G_2$ orthogonally transitive
perfect fluids}
\label{sect}

As stated in the introduction, we are interested in cosmological
abelian orthogonally transitive $G_2$ perfect-fluid 
solutions
of Einstein's field equations.
It is well-known that for such spacetimes
there
exist coordinates $\{ t, x, y,z \}$ adapted to the Killing
vectors in which the metric takes the form
\begin{eqnarray}
ds^2= \frac{1}{S^2(t,x)} \left [ \frac{}{} -dt^2 + dx^2 + F(t,x) \left (
P^{-1}(t,x) dy^2
+ P(t,x) \left ( dz + W(t,x) dy \right )^2 \right ) \right ].
\label{metgen}
\end{eqnarray}
The two Killing vectors are obviously given by
\begin{eqnarray*}
\vec{\xi} = \partial_y, \hspace{1cm}
\vec{\eta} = \partial_z.
\end{eqnarray*} 
We are going to analyze the Einstein field equations for the metric 
(\ref{metgen}) when the matter content corresponds to a perfect fluid.
Thus, the energy-momentum tensor takes the standard form
\begin{eqnarray*}
T_{\alpha\beta}= \left ( \rho + p \right )
u_\alpha u_\beta + p g_{\alpha\beta},
\end{eqnarray*}
where $\rho$ is the energy density of the fluid, $p$ is the pressure 
and $u_\alpha$ is the fluid 4-velocity. For orthogonally
transitive $G_2$ on $S_2$ spacetimes,
the fluid velocity vector is necessarily
orthogonal at each point to the group orbits.
Thus, the one-form $\mbox{\boldmath$u$}$ is
always a linear combination of the coordinate one-forms $\mbox{\boldmath$dt$}$
and
$\mbox{\boldmath$dx$}$ at each point and, consequently,
the fluid flow is irrotational. 
In all the cases below, we will 
write the Einstein field equations using 
orthonormal tetrads
$\{\mbox{\boldmath$\theta^{\alpha}$}\}$ such that 
\begin{eqnarray*}
\mbox{\boldmath$\theta^0$}= \frac{1}{S(t,x)}
\mbox{\boldmath$dt$}, \hspace{1cm} 
\mbox{\boldmath$\theta^1$} = \frac{1}{S(t,x)} \mbox{\boldmath$dx$}.
\end{eqnarray*}
Therefore, the fluid one-form will always take the form
\begin{eqnarray}
\mbox{\boldmath$u$}=u_0 \mbox{\boldmath$\theta^0$}+ u_1 \mbox{\boldmath$
\theta^1$}, \hspace{15mm} u_{0}^2-u_{1}^2=1,  \label{uu}
\end{eqnarray}
so that the Einstein field equations, in units where $c= 8 \pi G=1$, read
\begin{eqnarray*}
S_{00}  = \left(\rho + p\right ) u_0^2 - p, \hspace{1cm}
S_{01}  = \left ( \rho + p \right ) u_0 u_1, \hspace{1cm}
S_{11}  = \left ( \rho + p \right ) u_1^2 +p, \\
S_{22} = S_{33}  = p, \hspace{1cm}
S_{02}=S_{03} = S_{12}=S_{13}=S_{23} = 0, \hspace{2cm}
\end{eqnarray*}
where $S_{\alpha\beta}$ stands for the components of the Einstein tensor in the
$\{\mbox{\boldmath${\theta}^{\alpha}$}\}$ cobasis. 

Given that we are not assuming any particular equation of state for the
fluid, these Einstein field equations can be rewritten in terms of
the Einstein tensor only. The energy density, the
pressure and the form of the fluid velocity vector can be found
afterwards, once the field equations are solved.
Due to the separation into two orthogonal blocks
of the metric, some of the Einstein tensor components are identically
vanishing. In fact, a direct calculation
shows that 
$S_{02}$, $S_{03}$, $S_{12}$ and $S_{13}$ vanish
and consequently, the only equations we need to consider are
\begin{eqnarray}
\left ( S_{00}+S_{22} \right ) \left ( S_{11} - S_{22} \right ) -
S_{01}^2 = 0, \label{equ} \\
S_{22} - S_{33} =  0, \label{eqS2233} \\
S_{23} = 0. \label{eqS23}
\end{eqnarray}
Equations (\ref{equ})-(\ref{eqS23})  are satisfied by all perfect-fluid
metrics (with an orthogonally transitive $G_2$) but not all the 
solutions of this system represent a perfect-fluid spacetime.
Let us find which are the
restrictions we need to impose to obtain the perfect-fluid solutions
(in general these conditions can only be checked once the equations
are solved). The
quadratic equation (\ref{equ}) corresponds exactly to the vanishing
of the determinant of the $2 \times 2$ matrix
\[
\left ( \begin{array}{cc}
          S_{00}+S_{22}  & S_{01}\\
          S_{01} & S_{11} - S_{22}
         \end{array} \right),
\]
which is clearly equivalent to the existence of two functions $U_0$ and $U_1$
such that
\begin{eqnarray*}
S_{00}+S_{22} = \epsilon U_0^2, \hspace{1cm} S_{01}= \epsilon U_0 U_1,
\hspace{1cm} S_{11} - S_{22}= \epsilon U_1^2,
\end{eqnarray*}
where $\epsilon$ is an appropriate sign. Only when the two functions
$U_0$ and $U_1$ satisfy
\begin{eqnarray*}
U_0^2 - U_1^2 > 0  \label{cond1}
\end{eqnarray*}
we can define the two components $u_0$ and
$u_1$ of the fluid velocity vector through the expressions 
\begin{eqnarray*}
\epsilon U_0^2 \equiv \left (\rho + p \right) u_0^2,
\hspace{1cm}\epsilon U_1^2 \equiv \left (\rho + p \right) u_1^2,
\hspace{7mm} \mbox{with} \hspace{7mm} u_0^2-u_1^2=1,
\end{eqnarray*}
(which also define $\rho+p$ as $\rho+p = 
S_{00} - S_{11} + 2 S_{22}$). Obviously, for physically well-behaved
perfect fluids 
$\epsilon$ must be positive so that $\rho+p>0$, but this is not
necessary from a mathematical point of view for a solution of
(\ref{equ})-(\ref{eqS23}) to be a perfect fluid. 
The separate expression for
the density and pressure are obtained from
$S_{22}=S_{33}=p$. However, the solutions of (\ref{equ})-(\ref{eqS23})
can also satisfy $U_0^2 - U_1^2 \leq 0$.
When 
\begin{eqnarray*}
U_0^2 - U_1^2 = 0 \hspace{2mm} 
\Longleftrightarrow \hspace{2mm}  S_{00} + S_{22} = S_{11} - S_{22}
\end{eqnarray*}
the Einstein tensor (and therefore
the energy-momentum tensor) takes the form
\begin{eqnarray*}
S_{\alpha\beta}= K_{\alpha}K_{\beta} + p g_{\alpha\beta}
\end{eqnarray*}
where $\vec{K}$ is a null vector lying in the two plane spanned by the
vectors $\partial_t$ and $\partial_x$. This
kind of solution represents
a radiative fluid with pressure (which should satisfy $p=0$ to be 
physically reasonable). As we are looking for perfect-fluid cosmologies
we will not be interested in these radiative solutions
here.

Finally, when $U_0^2 - U_1^2 < 0$,
the energy-momentum corresponds to a perfect fluid but with
the ``fluid'' velocity being spacelike. 
We will obviously discard this case in the following. 

To summarize, the conditions that the solutions of the system
(\ref{equ})-(\ref{eqS23}) should satisfy in order 
to represent a perfect fluid can be written as
\begin{eqnarray}
S_{00} - S_{11} + 2 S_{22} \neq 0,\hspace{15mm}
\frac{S_{00}+S_{22}}{S_{00} - S_{11} + 2 S_{22}} > 0. \label{conditi}
\end{eqnarray}

Our next aim will be to impose the existence of a proper 
CKV in the spacetime (one of the main assumptions in
our work). Before entering into this we want to state
an interesting property for
general abelian orthogonally transitive $G_2$ on $S_2$ spacetimes
which will simplify considerably our work.
\begin{lemma}
If the spacetime
\begin{eqnarray}
ds^2 = \frac{1}{S^2(t,x)} \left [ -dt^2 + dx^2 + F(t,x) \left (
P^{-1}(t,x)
dy^2 + P(t,x) \left ( dz + W(t,x) dy \right )^2 \right ) \right ]
\label{prim}
\end{eqnarray}
satisfies the equations (\ref{equ}), (\ref{eqS2233}) and (\ref{eqS23}), then
the new metric
\begin{eqnarray}
ds^2 = \frac{1}{S^2(x,t)} \left [ -dt^2 + dx^2 + F(x,t) \left (
P^{-1}(x,t) dy^2 + P(x,t) \left ( dz + W(x,t) dy \right )^2 \right )
\right ] \label{sec}
\end{eqnarray}
obtained by interchanging $x$ and $t$ in all metric functions,
also satisfies the set of differential equations (\ref{equ}), (\ref{eqS2233})
and (\ref{eqS23}).
\end{lemma}
It is worth pointing out (and we will see that explicitly in the proof
of the lemma) that if the first spacetime represents a perfect fluid
then the second one is {\it not} a perfect-fluid solution because the
conditions (\ref{conditi}) are not satisfied (the sign of $u^2$ is
switched). Thus, this lemma
does not allow us to find new perfect-fluid solutions from old ones. The
interest of the result lies in the fact that the amount of work to be done
is reduced considerably in many problems. In particular, we will
see when imposing the existence of a CKV that
we need to consider only one half of all different possibilities.
Let us now  prove  the lemma.

{\it Proof:} In order to avoid confusion, we will write the second metric
(\ref{sec}) using coordinates $T$ and $X$ instead of $t$ and $x$. So the
metric reads
\begin{eqnarray}
ds^2 = \frac{1}{S^2(X,T)} \left [ -dT^2 + dX^2 + F(X,T) \left (
P^{-1}(X,T) dy^2 + \right . \right . \hspace{2cm} \nonumber \\
\;\;\;\left . \left . P(X,T) \left ( dz + W(X,T) dy \right )^2 \right )
\right ]. \label{sec2}
\end{eqnarray}
Let us now consider the following line-element
\begin{equation}
ds^2 = \frac{1}{S^2(t,x)} \left [ -\sigma dt^2 + \sigma dx^2 + F(t,x) \left (
P^{-1}(t,x)
dy^2 + P(t,x) \left ( dz + W(t,x) dy \right )^2 \right ) \right ],
\label{common}
\end{equation}
where $\sigma = \pm 1$. It is clear that when $\sigma =1$ this metric
is exactly the same as (\ref{prim}) while when $\sigma=-1$ the coordinate
change
\begin{eqnarray*}
t= X, \hspace{1cm} x= T
\end{eqnarray*}
transforms the metric into (\ref{sec2}).
We are now going to analyze the Einstein tensor for the metric (\ref{common}).
We choose the tetrad
\begin{eqnarray}
\mbox{\boldmath$\theta^0$} = \frac{1}{S(t,x)} \mbox{\boldmath$dt$},
 \hspace{3mm}
\mbox{\boldmath$\theta^1$} = \frac{1}{S(t,x)} \mbox{\boldmath$dx$},
 \hspace{2cm} \nonumber \\
\mbox{\boldmath$\theta^2$} = \frac{1}{S(t,x)} \sqrt{\frac{F(t,x)}{P(t,x)}}
\mbox{\boldmath$dy$}, \hspace{3mm}
\mbox{\boldmath$\theta^3$} = \frac{1}{S(t,x)} \sqrt{F(t,x) P(t,x)}
\left (\frac{}{}
\mbox{\boldmath$dz$} + W(t,x) \mbox{\boldmath$dy$} \right). \label{tetth}
\end{eqnarray}
The scalar products of these tetrad one-forms are, obviously,
\begin{eqnarray*}
\left ( \mbox{\boldmath$\theta^\alpha$}, \mbox{\boldmath$\theta^\beta$}
 \right ) = \mbox{diag} \left (- \sigma, \sigma, 1, 1 \right).
\end{eqnarray*}
The calculation of the Einstein tensor using this tetrad shows
that in changing $\sigma \rightarrow - \sigma$ the components
$S_{22}$, $S_{23}$, $S_{33}$ switch sign whereas $S_{00}$, $S_{01}$, $S_{11}$
remain unchanged
(all the rest of components of the Einstein tensor are zero in both
cases $\sigma=1$ and $\sigma=-1$). We must now compare the components
of the Einstein tensor of (\ref{sec2}) with the components of the
Einstein tensor of (\ref{prim}) when in both cases an orthonormal 
tetrad with timelike 
$\mbox{\boldmath$\theta^0$}$ is used (because this was used in 
(\ref{uu}) in order to write Einstein's equations in the form
(\ref{equ})-(\ref{eqS23})). When $\sigma=1$
the tetrad (\ref{tetth}) has already $\mbox{\boldmath$\theta^0$}$ timelike
but when
$\sigma=-1$ we must interchange the superscripts $0$ to $1$ because $x$ is
now the
timelike coordinate.
So, we find that the relationship between the Einstein tensors of the
metrics (\ref{prim}) and (\ref{sec}) using  respectively orthonormal
tetrads with $\mbox{\boldmath$\theta^0$}$ being timelike is given by
\begin{eqnarray*}
S_{00}^{(2)} = S_{11}^{(1)}, \hspace{1cm}
S_{01}^{(2)} = S_{01}^{(1)}, \hspace{1cm}
S_{11}^{(2)} = S_{00}^{(1)}, \\
S_{22}^{(2)} = - S_{22}^{(1)}, \hspace{1cm}
S_{23}^{(2)} = - S_{23}^{(1)}, \hspace{1cm}
S_{33}^{(2)} = - S_{33}^{(1)}, \label{relat}
\end{eqnarray*}
where the superscript $(1)$ denotes the first metric (\ref{prim}) and
the superscript $(2)$ denotes the second metric (\ref{sec2}). 
Consequently, if $S_{\alpha\beta}^{(1)}$ satisfies the equations
(\ref{equ}), (\ref{eqS2233}) and (\ref{eqS23}) then 
$S_{\alpha\beta}^{(2)}$ also satisfies the same equations and the lemma
is proven. Recalling that the conditions
for the solutions of these equations to represent perfect
fluids are (\ref{conditi})
and given that 
\begin{eqnarray*}
S_{00}^{(2)} + S_{22}^{(2)} = S_{11}^{(1)} - S_{22}^{(1)},
\hspace{1cm}
S_{11}^{(2)} - S_{22}^{(2)} = S_{00}^{(1)} + S_{22}^{(1)}, 
\end{eqnarray*}
we find that in spacetime regions where one metric represents a perfect
fluid, the other represents a tachyon fluid and vice versa.
\begin{flushright}
$\Box$
\end{flushright}
Until now the discussion has dealt with general abelian $G_2$ orthogonally
transitive perfect fluids. In the next section we will impose
the existence of a CKV, classify the inequivalent
Lie algebras and write down canonical line-elements
for each resulting case.

\section{Inequivalent Lie algebras and line-elements}

The main assumption in this paper is the existence of a proper conformal
motion in the manifold. We do not assume any further conformal symmetries
in the
manifold, but we assume that the two abelian KVs and the CKV,
$\vec{k}$, span a three-dimensional conformal Lie algebra
which, in general, satisfies
\begin{eqnarray}
\left[ \vec{\xi}, \vec{\eta} \right ] = \vec{0}, \hspace{5mm}
\left [ \vec{\xi}, \vec{k} \right ] = \gamma_1 \vec{\xi} + \gamma_2
\vec{\eta} + \gamma_3 \vec{k}, \hspace{5mm}
\left [ \vec{\eta}, \vec{k} \right ] = \delta_1 \vec{\xi} +
\delta_2 \vec{\eta} + \delta_3 \vec{k}, \label{GenLie}
\end{eqnarray}
where all $\gamma$'s and $\delta$'s are obviously constants (in the presence
of a fourth symmetry the existence of the three-dimensional Lie subalgebra
(\ref{GenLie}) could be violated). 
As a short calculation shows, all Lie algebras (\ref{GenLie}) are
solvable (see e.g. \cite{Pet} for a definition of solvability).
We in the following make a stronger
assumption. We assume that the two-dimensional
abelian subalgebra generated by the two KVs in (\ref{GenLie})
is an {\it ideal} (understanding the multiplication in the algebra as taking
the commutator), i.e.\ we assume $\gamma_3=\delta_3=0$. 
The line-element
and consequently the Einstein field equations take very different
forms whether the isometry subalgebra is an ideal or not. Thus, the two
cases must be treated separately.  In this paper we will concentrate only
on the case when the isometry subalgebra is an ideal.
{\it Thus, we assume from now on that $\gamma_3=
\delta_3=0$}. We must now find canonical forms for the inequivalent
Lie algebras contained
in (\ref{GenLie}) for $\gamma_3=\delta_3=0$. In order to do so, we can exploit
the freedom in performing linear transformations of the KVs
$\vec{\xi}$ and $\vec{\eta}$. Under these transformations, the $2 \times 2$
matrix
\begin{eqnarray*}
\left ( \begin{array}{cc}
        \gamma_1 & \gamma_2 \\
        \delta_1 & \delta_2
        \end{array}
\right )
\end{eqnarray*}
transforms like an endomorphism. By doing this,
we can reach a canonical Jordan form for this matrix. Three different possibilities arise depending on whether
the matrix diagonalizes with real eigenvalues (Case A), diagonalizes
with complex eigenvalues (Case VII) or does not diagonalize (Case B).
We used the term VII because that Lie algebra corresponds
to the Bianchi type VII in Bianchi's classification
of three-dimensional Lie algebras. Each one of the other two cases, A and B,
contains several different Bianchi types and no similar
Bianchi type name can be used (I,III,V,VI are in A and II,IV are in B). 
In the complex case, we then perform a transformation to obtain
a real valued matrix. The three possibilities read, finally,

\vspace{5mm}

{\bf Lie Algebra A}

\begin{eqnarray}
\left[ \vec{\xi}, \vec{\eta} \right ] = \vec{0}, \hspace{5mm}
\left [ \vec{\xi}, \vec{k} \right ] = \frac{1}{2} \left (c+b \right ) 
\vec{\xi}, \hspace{5mm}
\left [ \vec{\eta}, \vec{k} \right ] = \frac{1}{2} \left ( c - b \right ) 
\vec{\eta}, \label{LieA}
\end{eqnarray}
where $b$ and $c$ are arbitrary (possibly vanishing) constants.

\vspace{5mm}

{\bf Lie Algebra B}

\begin{eqnarray*}
\left [ \vec{\xi}, \vec{\eta} \right ] = \vec{0}, \hspace{5mm}
\left [\vec{\xi}, \vec{k} \right ] = \frac{1}{2} c \vec{\xi} + a \vec{\eta},
\hspace{5mm}
\left [ \vec{\eta}, \vec{k} \right ] = \frac{1}{2} c \vec{\eta},
\end{eqnarray*}
where $a$ is a non-vanishing constant.

\vspace{5mm}

{\bf Lie Algebra VII}

\begin{eqnarray*}
\left [ \vec{\xi}, \vec{\eta} \right ] = \vec{0}, \hspace{5mm}
\left [ \vec{\xi}, \vec{k} \right ] = \frac{1}{2} c \vec{\xi} - a
\vec{\eta},
\hspace{5mm}
\left [ \vec{\eta}, \vec{k} \right ] = a \vec{\xi} + \frac{1}{2} c 
\vec{\eta}.
\end{eqnarray*}
where $a$ and $c$ are constants ($a \neq 0$).

\vspace{5mm}

In order to find the line-elements for each of these Lie algebras we must
solve the conformal Killing equations 
\begin{eqnarray*}
{\cal L}_{\vec{k}} g_{\alpha\beta} = 2 \Phi g_{\alpha\beta},
\end{eqnarray*}
where ${\cal L}_{\vec{k}}$ represents the Lie derivative along the
vector field $\vec{k}$ and 
$\Phi$ is a scalar function usually called scale factor. Given that
we will concentrate on spacetimes admitting a {\it proper} CKV,
the scale factor will be assumed to be non-constant
(and in particular not identically zero). When solving the conformal
Killing equations, it is found that three different cases must
be considered depending on the character (spacelike, timelike
or null) of the orbits of the
three-dimensional conformal group (which is generated by the
Lie algebra $\{ \vec{\xi}, \vec{\eta}, \vec{k} \}$). In this paper
we will not study the case when these orbits are null and, instead, we
will concentrate on the cases when the orbits are either timelike or
spacelike. Let us first write down explicitly the canonical line-elements
for each of the Lie algebras A,B and VII when the orbits of the conformal
group are timelike.

\vspace{5mm}

{\bf Lie Algebra A}

\begin{eqnarray}
ds^2=\frac{1}{S^2(t,x)} \left[ -dt^2 + dx^2 + F(x) 
P^{-1}(x) e^{-(b+c)t} dy^2 + \right. \nonumber \\
\left. \frac{}{} + F(x) P(x) e^{(b-c)t} 
\left ( dz + W(x) e^{-bt} dy \right )^2
\right]. \label{metA}
\end{eqnarray} 
The expression for the conformal Killing vector in this metric is
\begin{eqnarray*}
\vec{k}= \partial_t + \frac{\left (c+b\right )}{2}y
\partial_y + \frac{\left(c-b\right)}{2} z \partial_z.
\end{eqnarray*}
The orthonormal tetrad we will use for this Lie algebra is
\begin{eqnarray}
\mbox{\boldmath$\theta^0$} = \frac{1}{S} \mbox{\boldmath$dt$}, \hspace{2mm}
\mbox{\boldmath$\theta^1$} = \frac{1}{S} \mbox{\boldmath$dx$}, \hspace{2mm}
\mbox{\boldmath$\theta^2$} = \frac{1}{S} \sqrt{\frac{F}{P}} e^{-\frac{b+c}{2}t}
\mbox{\boldmath$dy$}, \hspace{2mm}
\mbox{\boldmath$\theta^3$} = \frac{ \sqrt{FP}}{S} e^{\frac{b-c}{2} t} \left (
\mbox{\boldmath$dz$} + W e^{-bt} \mbox{\boldmath$dy$} \right). \label{tetA}
\end{eqnarray}

\vspace{5mm}

{\bf Lie Algebra B}

\begin{eqnarray}
ds^2 & = & \frac{1}{S^2(t,x)} 
\left[ \frac{}{} -dt^2 + dx^2 + \right. \nonumber \\
& & \;\;\;\;\;\;\;\;\;\;\;\; \left. \frac{}{} + F(x) e^{-ct} \left( 
P^{-1}(x)   dy^2 + P(x) \left( dz + \left[W(x) + a t
\right] dy \right)^2 \right) \right]. \label{metB}
\end{eqnarray}
The conformal Killing vector in this metric reads
\begin{eqnarray*}
\vec{k} = \partial_t + \frac{1}{2} c y 
\partial_y + \left (a y +\frac{1}{2} c z \right) \partial_z.
\end{eqnarray*}
The orthonormal tetrad we will use is
\begin{eqnarray}
\mbox{\boldmath$\theta^0$} = \frac{1}{S} \mbox{\boldmath$dt$}, \hspace{5mm}
\mbox{\boldmath$\theta^1$} = \frac{1}{S} \mbox{\boldmath$dx$}, \hspace{3cm}
\nonumber  \\
\mbox{\boldmath$\theta^2$} = \frac{1}{S} \sqrt{\frac{F}{P}}
e^{-\frac{c}{2} t } \mbox{\boldmath$ dy$}, \hspace{5mm}
\mbox{\boldmath $\theta^3$} = \frac{ \sqrt{FP}}{S} e^{-\frac{c}{2} t}
\left ( \mbox{\boldmath $dz$} + \left [ W + a t \right] \mbox{\boldmath $dy$}
\right ). \label{tetB}
\end{eqnarray}

\vspace{5mm}

{\bf Lie Algebra VII}

\vspace{5mm}

In this case the explicit form of the metric is much longer than before.
We will give, instead, the explicit form of the orthonormal
tetrad we will use in our calculations. It reads
\[
\mbox{\boldmath$\theta^0$} = \frac{1}{S(t,x)} \mbox{\boldmath$dt$},
\hspace{5mm}
\mbox{\boldmath$\theta^1$} = \frac{1}{S(t,x)} \mbox{\boldmath$dx$}, \]
\begin{equation}
 \mbox{\boldmath$\theta^2$} = \frac{1}{S(t,x)} \sqrt{\frac{F(x)}{P(x)}}
e^{- \frac{c}{2} t} \left [ \cos \left (a t \right) 
\mbox{\boldmath$dy$}
- \sin \left ( a t \right) \mbox{\boldmath$dz$} \right] ,
  \label{tetVII} 
\end{equation}
\[\mbox{\boldmath$\theta^3$} = \frac{\sqrt{ F(x) P(x)}
e^{- \frac{c}{2} t }}{S(x,t)}
 \left ( \frac{}{}  \left [ \cos \left (a t \right) - W(x)
\sin \left ( a t \right) \right ] \mbox{\boldmath$ dz$} + 
\left [ \sin \left ( a t \right) + W(x) \cos \left ( a t \right )
\right ] \mbox{\boldmath $dy$} \right).
\]
The conformal Killing vector in this metric is
\begin{eqnarray*}
\vec{k} = \partial_t + \left (\frac{1}{2} c y +
a z \right) \partial_y + \left (- a y + \frac{1}{2} c z
\right) \partial_z.
\end{eqnarray*}

The scale factor of the CKV for each one of these metrics
reads
\begin{eqnarray}
\Phi = - \frac{S_{,t}}{S}. \label{scafac}
\end{eqnarray}

For the remaining cases in which the orbits of the conformal group
are spacelike, the line-elements can be obtained
directly from the previous ones by simply interchanging $t$ and $x$ in the
metric coefficients as described in section \ref{sect}.
It is now obvious that making use of the lemma 1 we can restrict the
study to the case when the orbits of the conformal group are timelike
because in finding the general solution of the three equations (\ref{equ}),
(\ref{eqS2233}) and (\ref{eqS23}) we will also find the general solution
for the metrics obtained by interchanging the roles of $x$ and $t$ in
the metric coefficients, which correspond exactly to the metrics
when the orbits of the three-dimensional conformal group are
spacelike.

In the following section we will concentrate on the non-diagonal
metrics and will prove that there do not exist
perfect-fluid solutions under the assumptions of this paper when
the line-element is non-diagonal. 

\section{Non-diagonal metrics}

In this section we are going to prove that all the perfect-fluid
solutions of Einstein field equations for the spacetimes we are
considering in this paper (see beginning of section 3 for a clear account
of our assumptions) must be diagonal. In the previous
section we found the canonical line-elements for the three possible
Lie algebras (A, B and VII). It is clear from the expressions for these
line-elements that all the metrics belonging to Lie algebras B and VII
must be necessarily non-diagonal (irrespective of any field equations) and that
only for Lie algebra A the diagonal case is possible  (setting
$W=0$ in (\ref{metA})). Thus, in order to prove that no non-diagonal perfect
fluids exist, we must show that the metrics
in Lie algebras B and VII do not admit perfect-fluid solutions at all and that
only the diagonal metrics in Lie algebra A admit perfect-fluid
solutions.
These results will be stated in the following three propositions
(each one corresponding to one of
the Lie algebras A,B and VII).
Let us start with the simplest case, namely the
one corresponding to Lie algebra B.

\begin{proposition}
No perfect-fluid solutions exist for the conformal Lie Algebra B 
with $\vec{k}$ being a proper CKV.
\end{proposition}

{\it Proof:}
We will concentrate on the two Einstein field
equations (\ref{eqS2233}) and (\ref{eqS23}) which, using the tetrad 
(\ref{tetB}), read respectively
\begin{eqnarray}
\p \f + \Dp - \Qp - {\w}^2 P^2 + a^2 P^2 - 2 \p \frac{S_{,x}}{S} =0 
\label{S2233B} \\
\w \left ( \frac{1}{2} \f + \p - \frac{S_{,x}}{S} \right ) + \frac{1}{2}
W'' +\frac{1}{2} ac + a \frac{S_{,t}}{S} =0 \label{S23B}
\end{eqnarray}
where the prime denotes derivative with respect to $x$ and the comma
indicates partial derivative. Let us first consider the situation in which
$P' \neq 0$. The first equation above shows that the function $S$ must
take the form of a product of a function of $x$ and a function of $t$,
$S=G(x)H(t)$.
Then, the second equation reads
\begin{eqnarray*}
\w \left ( \frac{1}{2} \f + \p - \g \right ) + \frac{1}{2} W'' + \frac{1}{2}
ac + a \h =0,
\end{eqnarray*}
which immediately implies that $\dot{H}/H$ 
must be a constant (remember that for
the Lie algebra B the constant $a$ cannot vanish) and then
the expression for the scale factor (\ref{scafac}) implies that the
CKV is in fact homothetic. Thus no perfect fluid with a 
proper conformal motion does exist in this case. It only remains to
analyze the situation when $P$ satisfies
$P' \equiv 0$. The equation (\ref{S2233B}) gives simply
\begin{eqnarray*}
\w = \epsilon_1 a,
\end{eqnarray*}
where $\epsilon_1 = \pm 1$. The second equation (\ref{S23B}) takes the form
\begin{eqnarray*}
\epsilon_1 \frac{1}{2} \f + \frac{1}{2} c + \frac{S_{,t}}{S} - \epsilon_1
\frac{S_{,x}}{S} =0,
\end{eqnarray*}
which can be integrated to give
\begin{eqnarray*}
S = \sqrt{F} \exp \left (\epsilon_1 \frac{c}{2} x\right)
U \left (t+\epsilon_1 x \right )
\end{eqnarray*}
with $U$ an arbitrary function of $t+ \epsilon x$. The quadratic
Einstein equation (\ref{equ}) takes now the very simple form
\begin{eqnarray*}
\left (\Df - \Qf \right )^2=0 \hspace{5mm}
\Longleftrightarrow \hspace{5mm} F= F_0 e^{\beta x},
\end{eqnarray*}
with constants $F_0$ and $\beta$. Now we can evaluate the components
of the Einstein tensor to find
\begin{eqnarray*}
S_{00} = S_{11} = \epsilon_1 S_{01}, \hspace{2cm} S_{22} = S_{33} = S_{23} = 0,
\end{eqnarray*}
which shows that the energy-momentum content is that of a null fluid without
pressure. Thus, we can conclude that no perfect-fluid solutions with 
a proper CKV exist for the Lie algebra of the type B.
Let us emphasize once more that this analysis covers both cases,
when the orbits of the conformal group are timelike and spacelike.

Let us now prove a similar result for Lie algebra VII, i.e.\ that no
perfect-fluid solutions arise for Lie algebra VII.

\begin{proposition}
No perfect-fluid solutions of Einstein's field equations do exist for
spacetimes
\begin{itemize}
\item[-] admitting a three-dimensional Lie algebra of class VII of two
KVs and one proper CKV with
\item[-] the two-dimensional isometry group being maximal.
\end{itemize}
\end{proposition}

{\it Proof:} The proof for this case is similar to the previous one and
we will only outline the main steps. We concentrate, as before, on
the two Einstein equations (\ref{eqS2233}) and (\ref{eqS23}) which,
using the tetrad (\ref{tetVII}), read respectively
\begin{eqnarray}
\f \p + \left ( \p \right )' - 2 \p \frac{S_{,x}}{S} -
4 a W \frac{S_{,t}}{S} - \hspace{3cm} \nonumber \\
- {\w}^2 P^2 + {a}^2 P^2 \left (W^2 +1  
\right )^2 - \frac{a^2}{P^2} - 2 a c W = 0 \label{S2233VII}\\
\w \left ( \frac{1}{2} \f + \p - \frac{S_{,x}}{S} \right) + a 
\left ( \frac{S_{,t}}{S} + \frac{c}{2} \right ) \left ( W^2 +1 -
\frac{1}{P^2} \right ) + \hspace{1cm} \nonumber \\
+ \frac{1}{2} W'' + a^2 W \left ( W^2 +1 +
\frac{1}{P^2} \right )=0. \label{S23VII}
\end{eqnarray}
We must again distinguish between two cases depending on whether $P' 
\equiv 0$ or not. When $P=\mbox{const}$, the only non-homothetic solution
of the system (\ref{S2233VII})-(\ref{S23VII}) is $P=1$ and $W=0$. Then,
the tetrad reduces to
\begin{eqnarray*}
\mbox{\boldmath $\theta^0$} = \frac{\mbox{\boldmath $dt$}}{S},
\hspace{5mm} \mbox{\boldmath $\theta^1$} = \frac{\mbox{\boldmath $dx$}}{S}, 
\hspace{5mm} 
\mbox{\boldmath $\theta^3$} = \frac{1}{S} \sqrt{F}
e^{- \frac{c}{2} t} \left ( \cos \left (a t \right)
\mbox{\boldmath $ dy$}
-\sin \left (a t \right) \mbox{\boldmath $dz$ } \right ), \\
\mbox{\boldmath $\theta^4$} = \frac{1}{S} \sqrt{F}
e^{- \frac{c}{2} t} \left (\sin \left (a t \right)
\mbox{\boldmath $dy$}
+ \cos \left ( a t \right ) \mbox{\boldmath $dz$ } \right ) \hspace{3cm}
\end{eqnarray*}
and the line-element is simply
\begin{eqnarray*}
ds^2 = \frac{1}{S(x,t)^2} \left [ -dt^2 + dx^2 + F(x) e^{-c t}
 \left ( dy^2 + dz^2
\right ) \right ]
\end{eqnarray*}
which has a three-dimensional isometry group (against our assumptions). Thus,
we must consider the case $P' \neq 0$. It is not difficult to see
that the only possibility for proper conformal solutions is 
\begin{eqnarray*}
S = G(x) H(t + a K(x)), \hspace{3mm} \left(\mbox{with}
\hspace{5mm} K' = - 2 W \frac{P}{P'}\right) 
\\
\p = 2 \epsilon_2 a W, \hspace{1cm} \g = \frac{1}{2}
 \left ( \f - \epsilon_2
c \right ), \hspace{1cm} W^2 = \frac{m^2}{P} - 1 - \frac{1}{P^2},
\end{eqnarray*} 
where $m$ is a constant, 
$\epsilon_2 = \pm 1$ and $H$ is an arbitrary function of its argument.
Imposing now the quadratic
Einstein equation (\ref{equ}) we simply find
\begin{eqnarray*}
\left ( \Df - \Qf \right )^2 =0 \hspace{5mm}\Longleftrightarrow 
\hspace{5mm} F = F_0 e^{\beta x },
\end{eqnarray*}
where $F_0$ and $\beta$ are constants of integration.
As in the previous case, the evaluation of the Einstein tensor for this
solution gives us
\begin{eqnarray*}
S_{00} = S_{11} = - \epsilon_2 S_{01}, \hspace{1cm} S_{22} = S_{33} = S_{23} =0
\end{eqnarray*} 
so that the matter content is a null fluid without pressure and
the proof of the proposition is completed.

Only Lie algebra A remains to be analyzed. Let us prove
that all the perfect-fluid solutions with a proper CKV belonging to Lie
algebra A must be diagonal.

\begin{proposition}
No perfect-fluid solutions of Einsteins field equations do exist for
spacetimes 
\begin{itemize}
\item[-] admitting a three-dimensional Lie algebra of class A of two Killing
vectors and one proper CKV with
\item[-] the maximal two-dimensional isometry group being maximal and
\item[-] the metric being non-diagonal.
\end{itemize}
\end{proposition}

The proof is again similar and we will just sketch it. The
two equations 
(\ref{eqS2233}) and (\ref{eqS23}) read respectively
\begin{eqnarray}
\p  \f  + \left ( \p \right )' - 2 \p \frac{S_{,x}}{S} + 2 b \frac{S_{,t}}{S}
- {W'}^2 P^2 + b^2 P^2 W^2 + b c = 0 , \label{eqS2233A} \\
W' \left ( \frac{1}{2} \f + \p - \frac{S_{,x}}{S} \right ) + \frac{1}{2} W''
+ \frac{1}{2} b \left (b-c \right ) W - b W \frac{S_{,t}}{S} =0 . 
\label{eqS23A}
\end{eqnarray}
The $P' \equiv 0$ subcase gives rise either to homothetic spacetimes or to
metrics 
with a three-dimensional group
of isometries acting on two-dimensional spacelike surfaces.
Thus, we can move
to the general case $P' \neq 0$. Equation (\ref{eqS2233A}) gives
\begin{eqnarray*}
S= G(x) H(t + b K(x)) \hspace{7mm} \mbox{with} \hspace{5mm} K' = \frac{P}{P'}.
\end{eqnarray*}
which inserted into (\ref{eqS23A}) produces
\begin{eqnarray*}
W' \left ( \frac{1}{2} \f + \p - \g \right ) + \frac{1}{2} W'' + \frac{1}{2} 
b \left (b-c \right ) W = \h b \left ( W + W' \frac{P}{P'} \right ).
\end{eqnarray*}
The conditions that the CKV is proper implies
that both sides in this equation must vanish. In particular, the
right-hand side implies that either $b=0$ or $W= \alpha/P$ (where 
$\alpha$ is
an arbitrary constant). Let us start with $b=0$. The
two equations (\ref{eqS2233A}) and (\ref{eqS23A}) take now the form
\begin{eqnarray} 
\p \left ( \f - 2 \g \right ) + \left ( \p \right )' - {W'}^2 P^2 =0, 
\label{uno}\\
W' \left ( \frac{1}{2} \f - \g + \p \right ) + \frac{1}{2} W'' =0 \label{dos}
\end{eqnarray}
and the second one can be integrated to give
\begin{eqnarray*}
W' = W_0 \frac{G^2}{F P^2 },
\end{eqnarray*}
where $W_0$ is an arbitrary constant. The case $W_0 =0 $ gives a diagonal
metric and, therefore,
we can assume $W_0 \neq 0$. Then,  equation 
(\ref{uno}) can be integrated to give
\begin{eqnarray}
\hspace{1cm} \frac{1}{P^2} = \delta^2 - \left (W + \beta \right )^2
\label{nondiag}
\end{eqnarray}
(where $\beta$ and $\delta$ are arbitrary constants of
integration). The metric is apparently non-diagonal, but
a trivial calculation using (\ref{nondiag}) shows that the
linear coordinate change in the block $\{y,z \}$
\begin{eqnarray*}
y = Y  + Z , \hspace{5mm}
z = \left ( \beta + \delta \right)
Y + \left (\beta - \delta \right ) Z 
\end{eqnarray*}
brings the metric into a diagonal form and therefore no proper non-diagonal
solutions do exist in this subcase (notice that this coordinate change
is singular only when $\delta = 0$ which is impossible due to
(\ref{nondiag})). 

We can now consider that last possibility, namely, when
$b\neq 0$ and
$W= \alpha/P$ (with  $\alpha \neq 0 $ in order to 
consider non-diagonal metrics). Substituting $W$ everywhere, it is easy
to see that the general solution of 
the full set of Einstein field equations
(\ref{equ}), (\ref{eqS2233}) and (\ref{eqS23}) is
\begin{eqnarray*}
\p = \sigma b, \hspace{1cm} \g = \frac{1}{2} \left (n + \sigma c
\right ), \hspace{1cm} \f = n,
\end{eqnarray*}
where $\sigma= \pm 1$ and $n$ is an arbitrary constant.
As in the two previous cases this solution
corresponds to a null fluid and the proof of the proposition is completed.

To summarize, we have proven in this section that
there do not exist any non-diagonal perfect-fluid solutions for
spacetimes with 
\begin{itemize}
\item[-] a maximal abelian
orthogonally transitive $G_2$ on $S_2$
\item[-] admitting one proper CKV (with
conformal orbits either 
spacelike or timelike) such that
\item[-]
the subalgebra generated by the two Killing
vectors is an ideal in the Lie algebra spanned by the two KVs and the CKV.
\end{itemize}
In the following section we will concentrate
on the diagonal case and find all the perfect-fluid
solutions (with $b\neq 0$, see the introduction). 

\section{Diagonal perfect-fluid solutions}
 
We know that only the Lie algebra A admits diagonal metrics (all metrics
in Lie algebras B and VII are necessarily non-diagonal and, as we have seen
in the previous section, none of them satisfies the perfect-fluid Einstein
field equations). 
In order to restrict the metrics for Lie algebra A
to be diagonal we must set
$W \equiv 0$ in (\ref{metA}) (and also in the corresponding tetrad
(\ref{tetA})). Now, the calculation of the Einstein tensor in this
tetrad gives 
$S_{23} \equiv 0$ so that the Einstein field
equation (\ref{eqS23}) is identically satisfied (this can be seen directly
from ({\ref{eqS23A}) after substituting $W=0$). Equation (\ref{eqS2233A})
with $W=0$ gives ($P= \mbox{const.}$ would imply $b=0$ and 
a third KV):
\begin{eqnarray}
S = G(x) H(t+b K(x)) \hspace{1cm} \mbox{with} \hspace{1cm} K' = 
\frac{P}{P'}, \nonumber  \\
\p \left ( \f - 2 \g \right ) + \left (\p \right )' + bc =0.
\label{qu2233}
\end{eqnarray} 
Now, we must inevitably analyze the more difficult quadratic equation
(\ref{equ}) which takes the form
\begin{eqnarray}
\Sigma_1 (x) + \h \Sigma_2(x) + \frac{\dot{H}^2}{H^2} \Sigma_3(x) +
\frac{\ddot{H}}{H} \frac{\dot{H}}{H} \Sigma_4(x) + \frac{\ddot{H}}{H}
\Sigma_5(x) =0,
\end{eqnarray}
where $\Sigma_i(x)$ ($i=1$ to $5$) are expressions 
containing $P$, $F$ and $G$ and their derivatives and which
depend only on the variable $x$. The dot means from now on 
derivative with respect
to the variable $t+ b K(x)$. In the Appendix
we find all the different
possibilities compatible with this 
mixed equation  containing both independent variables $x$ and
$t+ bK$. Let us here just summarize the results.

\vspace{5mm}

{\bf Case 1}

\begin{eqnarray*}
\Sigma_1= \Sigma_2 = \Sigma_3 = \Sigma_4 = \Sigma_5 = 0, \hspace{1cm}
H \hspace{5mm} \mbox{arbitrary}
\end{eqnarray*}

\vspace{5mm}

{\bf Case 2}
\begin{eqnarray*}
\Sigma_4 =0 , \hspace{5mm} \Sigma_1 = m_1 \Sigma_5, \hspace{5mm}
\Sigma_2= m_2 \Sigma_5, \hspace{5mm} \Sigma_3= m_3 \Sigma_5, \hspace{5mm}
\Sigma_5 \neq 0 \\
m_1 + m_2 \h + m_3 \frac{\dot{H}^2}{H^2} + \frac{\ddot{H}}{H} =0, \hspace{2cm}
\end{eqnarray*}
where $m_1$, $m_2$ and $m_3$ are arbitrary constants.

\vspace{5mm}

{\bf Case 3}

\begin{eqnarray*}
\Sigma_1 = k_1 \Sigma_4, \hspace{5mm} \Sigma_2= k_2 \Sigma_4, \hspace{5mm}
\Sigma_3 = k_3 \Sigma_4, \hspace{5mm} \Sigma_5 = k_5 \Sigma_4,
\hspace{5mm} \Sigma_4 \neq 0 \\
k_1 + k_2 \h + k_3 \frac{\dot{H}^2}{H^2} + \h \frac{\ddot{H}}{H} +
k_5 \frac{\ddot{H}}{H} = 0, \hspace{2cm}
\end{eqnarray*}
where $k_1$, $k_2$, $k_3$ and $k_5$ are arbitrary constants.

\vspace{5mm}

{\bf Case 4}
\begin{eqnarray*}
\Sigma_1 = k_1 \Sigma_5, \hspace{5mm} \Sigma_2 = k_1 \Sigma_4 + k_3
\Sigma_5, \hspace{5mm} \Sigma_3 = k_3 \Sigma_4, \hspace{5mm}
 \Sigma_5 \neq 0, \hspace{5mm} \frac{\Sigma_4}{\Sigma_5} \neq
\mbox{const.} \\
k_1 + k_3 \h + \frac{\ddot{H}}{H} =0 \hspace{4cm}
\end{eqnarray*}
where $k_1$ and $k_3$ are arbitrary constants.

The explicit expressions for $\Sigma_i$ will be necessary later to
study these four different cases. They can be directly obtained
from the following expressions.
\begin{eqnarray*}
\Sigma_1 \equiv Z_1 V_1 - W_1^2, \hspace{1cm}
\Sigma_2 \equiv Z_1 V_2 + Z_2 V_1 - 2 W_1 W_2, \hspace{1cm}
\Sigma_3 \equiv Z_2 V_2 - W_2^2, \\
\Sigma_4 \equiv -2 \left ( 2 W_2 R + Z_2 R^2 + V_2 \right ),
\hspace{1cm} \Sigma_5 \equiv -2 \left (2 W_1 R + Z_1 R^2 + V_1
\right ), \hspace{5mm}
\end{eqnarray*}
where $Z_1$, $Z_2$, $V_1$, $V_2$, $W_1$, $W_2$ and $R$ are given
explicitly in terms of the metric coefficients by
\[
Z_1 \equiv \frac{1}{2} \Df - \f \g + \frac{1}{2} b^2, \hspace{1cm}
Z_2 \equiv -c - b \f \frac{P}{P'},  \]
\[
V_1 \equiv \frac{1}{2} \Df - \frac{1}{2} \Qf + \f \g - 2 \Dg
+\frac{1}{2} \Qp + \frac{1}{2} c^2, \hspace{1cm}
V_2 \equiv c - 2 b^2 c \frac{P^2}{{P'}^2} - b \f \frac{P}{P'}, \]
\[
W_1 \equiv \frac{c}{2} \f - \frac{b}{2} \p, \hspace{1cm}
W_2 \equiv 2 \g, \hspace{1cm} R \equiv b \frac{P}{P'}. \]
As explained in the introduction, 
the case when the constant $b$ in the Lie algebra A is identically
zero has been exhaustively analyzed in \cite{COA}. Thus, we can
restrict our study to the case when $b \neq 0$. We have to solve
the four different sets of differential equations 
Case 1 to Case 4. Obviously, the two first cases are simpler
because of the equation $\Sigma_4 =0$. Let us first
concentrate on these two cases. It is convenient to define three
new functions $q(x)$, $l(x)$ and $m(x)$ through the
relations
\begin{eqnarray*}
\p \equiv b q, \hspace{1cm} \f \equiv \left ( c+ l \right ) q,
\hspace{1cm} \g \equiv \frac{1}{q} \left (c + \frac{m}{4} \right).
\end{eqnarray*}
The equation (\ref{qu2233}) reads now
\begin{eqnarray*}
q' + \left ( c + l \right) q^2 - c - \frac{m}{2} =0
\end{eqnarray*}
which allows us to obtain $m$ in terms of $q$ and $l$. Using this expression
for $m$, the equation $\Sigma_4 =0$ takes the form
\begin{eqnarray}
-2 q' + \left (l + 2c \right ) \left (1 - q^2 \right )  = 0. 
\label{sigma4}
\end{eqnarray}
The solution $q \equiv \pm 1$ of this equation also implies
$\Sigma_5 \equiv 0$ and, therefore, the
only possibility is Case 1. The expressions for 
$\Sigma_2$ and $\Sigma_3$ are identically 
zero while the vanishing of $\Sigma_1$ imposes 
$l'=0$. Now all the equations are satisfied. However, evaluating the Einstein
tensor for this solution we find
\begin{eqnarray*}
S_{00} + S_{22} = S_{11}- S_{22} = \pm S_{01}, \hspace{1cm}
S_{22} = S_{33} = 0
\end{eqnarray*}
so that it does not represent a perfect fluid. Thus, we can assume
$q^2 \neq 1$ and obtain $l$ from (\ref{sigma4}) as
\begin{eqnarray*}
l = -2c - \frac{2q'}{q^2 -1}.
\end{eqnarray*}
The expressions for $\Sigma_5$ and $\Sigma_3$ read now, respectively,
\begin{eqnarray*}
\Sigma_5 = - \frac{1}{q^2} \left [ \left ( q' + c q^2 -c \right)^2
+ \left (b^2 - c^2 \right) \left (q^2 -1 \right )^2 \right ], \\
\Sigma_3 = - \frac{1}{q^2} \left (q' +c q^2 - c \right )^2. \hspace{2cm}
\end{eqnarray*} 
Let us start by analyzing Case 1. Given that both $\Sigma_5$ and
$\Sigma_3$ must vanish we immediately find 
\begin{eqnarray*}
q' = c \left (1 - q^2 \right), \hspace{1cm} b = \epsilon c,
\end{eqnarray*}
where $\epsilon$ is a sign. 
Now, all $\Sigma_i$ are identically zero and we have another
solution of the Einstein field equations. 
The
calculation of the Einstein tensor shows that the matter content is
indeed a perfect fluid. However, the spacetime is conformally flat
and both the density and pressure depend only
on the variable $t + b K$. Thus they satisfy an equation of state 
$p=p(\rho)$ and,
consequently, the spacetime is a Friedman-Lema\^{\i}tre-Robertson-Walker
cosmology. Consequently, Case 1 does not give new perfect-fluid
solutions and we can move on to Case 2.

The equation $\Sigma_3 - m_3 \Sigma_5=0$ reads 
\begin{eqnarray}
\left (1 - m_3 \right ) \left ( q' + cq^2 - c \right)^2 
= m_3 \left (b^2 - c^2 \right ) \left (q^2 -1 \right)^2. \label{Eqs23}
\end{eqnarray}
When $m_3 \neq 1$ this equation immediately implies
\begin{eqnarray}
q' = n \left (1 - q^2  \right), \label{kpr}
\end{eqnarray}
where $n$ is a constant. When $m_3 =1$ the equation (\ref{Eqs23}) implies
$b = \epsilon c$ and then, combining the other two differential
equations $\Sigma_1 - m_1 \Sigma_5=0$ and $\Sigma_2 - m_2 \Sigma=0$ 
(two equations for one unknown $q(x)$), it is not difficult to show that
the relation (\ref{kpr}) must still hold. Substituting this
expression for $q'$ into all $\Sigma_i$ we find that 
all the equations of Case 2 are
fulfilled, and they simply give $m_1$, $m_2$ and $m_3$ in terms
of the constants $b$, $c$ and $n$. The explicit expressions are
\begin{eqnarray*}
m_3 = \frac{ \left (c-n \right)^2}{n^2 - 2cn + b^2},
\hspace{1cm} m_2 = \frac{ \left (n-c \right)^3 + 3c \left (n-c \right)^2
+ c \left (b^2 - c^2 \right) }{n^2 - 2cn + b^2},\\
 m_1 = \frac{1}{4} \frac{ \left (c^2 - n^2 \right)
\left( c^2 - 2cn + b^2 \right)}{n^2 - 2cn + b^2} \hspace{25mm}
\end{eqnarray*}
and the metric functions are given by
\[
\p = b q(x), \hspace{3mm} \f = \left (2n -c \right) q(x), \hspace{3mm}
\g = \frac{1}{2} \left (n- c \right) q(x) + \frac{1}{2} \frac{n + c}{q(x)}, 
\hspace{3mm} q' = n \left (1 - q^2 \right),
\]
where $b$, $c$ and $n$ are arbitrary constants.
Let us label this solution as {\bf Solution A}. We
will write down the explicit line-element and analyze the physical
content of this solution in the following section. 

Thus, we have exhausted
the Cases 1 and 2 and we have to deal now with the much more complicated
Cases 3 and 4. We have to take advantage that in both cases the
equation $\Sigma_3 - k_3 \Sigma_4 =0 $ holds. Therefore, we will try to write
this equation in the simplest possible form so that the
problem becomes more tractable. To that aim let us define three functions
$r(x)$, $u(x)$ and $v(x)$ through the relations
\begin{eqnarray*}
\f = 2 r(x) + u(x), \hspace{1cm} \g = r(x), \hspace{1cm}
\p = \kappa b \sqrt{v(x)},
\end{eqnarray*}
where $\kappa$ is the sign of $P'/(b P)$ (so that the square root is
non-negative definite, as usual).
Now, equation (\ref{qu2233}) allows us to obtain $u$
in terms of $v$ as 
\begin{eqnarray*}
u = - \frac{2c \kappa \sqrt{v} + v'}{2v}
\end{eqnarray*}
and substitute it in all other expressions so that only the two
functions $v(x)$ and $r(x)$ remain in the equations.

Let us consider the situation when $v$ is a constant. Then we must assume
$v \neq 1$ because otherwise we would have $\Sigma_4 \equiv 0 $ and would
be in a previous case. Now, the equation $\Sigma_3 - k_3 \Sigma_4 = 0$
immediately implies $r = \mbox{const.}$ So, we have 
\begin{eqnarray*}
\f = 2 a_1 - \frac{c}{a_2}, \hspace{1cm} \g = a_1, \hspace{1cm}
\p = b a_2
\end{eqnarray*}
($a_1$ and  $a_2$ constants)
and we certainly have a solution of Case 3 (Case 4 is impossible in
this situation because $\Sigma_5$ is constant and therefore proportional
to $\Sigma_4$). The equations simply give $k_1$, $k_2$, $k_3$ and $k_5$
in terms of the constants $c$, $b$, $a_1$ and $a_2$. This is a solution
of the Einstein field equations and its matter content is a perfect
fluid, but it is trivial to checked that the metric contains a homothetic
Killing vector (besides the proper CKV) and
therefore does not belong to the class we are interested in this paper.

So, we can assume from now on that $v$ is non-constant and 
perform the following
change of variables
\begin{eqnarray*}
dv = 2 \kappa \sqrt{v} \left ( v -1 \right) \left [\chi(v) - k_3
\right] dx
\end{eqnarray*} 
so that now the independent variable is $v$ and $\chi(v)$ is the
unknown. Let us also define a
function $\omega(v)$ through
\begin{eqnarray*}
r \equiv \frac{1}{2} \frac{\kappa}{\sqrt{v}} \left (\omega - \chi + 2 k_3
\right)
\end{eqnarray*}
so that we replace $r$ by $\omega$ everywhere.
With all these changes and redefinitions, the common equation
$\Sigma_3 - k_3 \Sigma_4 =0 $ takes the simple form
\begin{eqnarray}
v \left [ \chi^2 - \left (c - k_3 \right)^2 \right ] = \omega^2 -
\left (c - k_3 \right)^2. \label{s3s4}
\end{eqnarray}
This equation is linear in the independent variable $v$ and, therefore, it
would be convenient to consider $\chi$ as the independent variable
and $v(\chi)$ the unknown. This change of variables 
can be done unless $\chi$ is  constant.
Thus, we must consider the two cases $\chi=$ const. and $\chi$
non constant separately. Let us
first assume $\chi = \mbox{const.}$

Now, we have only one
unknown $\omega(v)$ which must satisfy an overdetermined system of
differential equations (four for Case 3 and three for Case 4). 
It is not difficult to
find the solutions which are:

\vspace{5mm}

{\bf Case 3 and $\mbox{\boldmath$\chi =$}$ const.}

\vspace{5mm}

The solution in this case is 
\begin{eqnarray*}
\chi = k_3 - \sigma_1 b , \hspace{1cm} \omega = k_3 - \sigma_1 b,
\hspace{1cm} c= \sigma_1 b, \\
k_1 = -\frac{3 b^2}{4}
\left (\sigma_1 b - k_3 \right ), \hspace{1cm} k_2= \frac{b}{2}
\left ( -b + 4 \sigma_1 k_3 \right), 
\hspace{1cm} k_5 = \frac{3}{2} \sigma_1 b, 
\end{eqnarray*}
where $\sigma_1 = \pm 1$ and $b$ and $k_3$ are arbitrary constants. Let us label this solution as {\bf Solution B}
for further study in the next section.
\vspace{5mm}

{\bf Case 4 and $\mbox{\boldmath$\chi =$}$ const.}

\vspace{5mm}

Now, two different solutions are possible. The first one is
\[
\chi = c - k_3, \hspace{9mm} \omega = c- k_3, \hspace{9mm} k_1 =0,
\hspace{9mm} b^2 + c^2 -2 c k_3 =0, \hspace{9mm} 
\mbox{$b, c$ arbitrary}
\]
which again will be studied in the following section. Let us call it
{\bf Solution C}. The other possibility reads
\begin{eqnarray*}
\omega = \sigma_2 \left ( c - \sigma_1 b \right) \sqrt{v} 
, \hspace{1cm} p= c - \sigma_1 b, \hspace{1cm} k_3 = c, \hspace{1cm}
k_1 = \frac{c^2 - b^2}{4},
\end{eqnarray*}
where $\sigma_2$ is also a sign. This solution, however,
can be seen to represent a null fluid (with
non-zero pressure, in general) and,
therefore, will not be analyzed any further. 

Having written all the solutions of the equations when $\chi=$ const,
let us consider the more general case when $\chi$ is non-constant.
Now, as stated above, it is convenient to perform another change
of variables and consider $\chi$ as
the independent variable and $v$ and $\omega$ as unknown functions
depending on $\chi$. Equation $\Sigma_3 - k_3 \Sigma_4 = 0$, 
(\ref{s3s4}), allows us to obtain $v$ in terms of $\chi$ as
\begin{eqnarray*}
v = \frac{\omega^2 - \left (c - k_3 \right)^2}{\chi^2 - \left (c- k_3
\right)^2 }
\end{eqnarray*}
and to substitute this expression in all the $\Sigma_i$. Now we have only
one unknown $\omega(\chi)$ which must satisfy an overdetermined
system of differential equations (three differential equations for 
Case 3 and two for Case 4). The investigation of the
compatibility of these systems
of differential equations is now rather lengthy  and makes essential
use of computer algebra \cite{FiHe}. Special care is needed to combine the
equations in the right order to avoid too long expressions.
Let us here simply summarize which are the results obtained. It turns
out that there exists one family of solutions for Case 3 and another
family for Case 4. They are, explicitly,

\vspace{5mm}

{\bf Case 3 and $\mbox{\boldmath$\chi$}$ non constant}
\begin{eqnarray*}
\omega = \chi - 2 \sigma_1 b +2 k_3, \hspace{1cm} c= \sigma_1 b,
\hspace{1cm} k_1 =0, \\
k_2 = \frac{b}{2} \left (
2 \sigma_1 k_3 - b \right), \hspace{1cm} k_5 = k_3 - \frac{\sigma_1 b}{2},
\hspace{1cm} \mbox{$b$ and $k_3$ arbitrary}.
\end{eqnarray*}
This solution will be labeled as {\bf Solution D} and will be
studied in the following section.

\vspace{5mm}

{\bf Case 4 and $\mbox{\boldmath $\chi$}$ non constant}

\begin{eqnarray*}
\omega = \chi \left (1 + \sigma_2 \sqrt{ \frac{k_3^2 - 4 k_1 - b^2}{
\left ( \chi - k_3 + b \right) \left (\chi- k_3 - b \right)}} 
\right) \hspace{1cm}
\mbox{ $k_1$ and $k_3$ arbitrary}. 
 \end{eqnarray*}
The Einstein tensor for this solution satisfies
$ S_{00}+S_{22} = S_{11}-S_{22} = - S_{01}$
and, therefore, as discussed above, represents a null fluid.
Consequently, it will not be analyzed any further.

To summarize, we have found four families of solutions which represent
perfect fluids when the spacetime is assumed to have an
abelian orthogonally transitive $G_2$ on $S_2$
and also one proper CKV which spans
together with the two KVs a Lie algebra
generating either timelike or spacelike orbits such that the isometry
subalgebra is an ideal. These
families have been labeled as  Solution A, B, C and D respectively and
will be studied in the next section.

\section{Analysis of the Perfect-Fluid Solutions}

The aim of this section is to write down the line-elements for the four
solutions found in the previous section (Solutions A, B, C and D). In order to
do so, we must integrate back some changes of variables which helped us
to solve the Einstein field equations.
We will not give the details as the calculations
are not difficult and can be easily reproduced.
In some cases we have also performed some redefinitions of constants
and some coordinate
changes in the block $\{t,x\}$ as well as rescalings of the coordinates
$y$ and $z$ in order to write the metrics in a
simplified form. Therefore constant names used in this section are
unrelated to previous appearances in earlier sections.
After giving explicitly the line-element (no differential equations
remain to be solved) 
we will study some of their properties.
In particular, we will
give the Petrov type, the
fluid 4-velocity, the kinematical quantities, the energy-density and the
pressure. 

We will also study in which spacetime region the solution
represents a perfect fluid. As discussed earlier, there is a
symmetry between exchanging $x$ and $t$ in the metric coefficients and
switching between a timelike perfect fluid and a tachyonic fluid.
In the case of metrics A and D there are regions of both, perfect and
tachyonic fluid. Here a switch $x \leftrightarrow t$ might be of
interest to exchange the physically meaningful regions.

Further, we describe for which ranges of the parameters the spacetime
satisfies energy conditions everywhere (in the perfect-fluid region)
and whether the fluid satisfies an equation of state $p=p(\rho)$ or not.

A brief description of the curvature singularities of the spacetime and 
an identification of the spatially homogeneous
and isotropic limits in each family
are also given. All four families have the magnetic part of the Weyl tensor
with respect to the fluid 4-velocity being non-zero.

Tensor components have been calculated with the help of
the computer program
CLASSI \cite{MSMM} using the orthonormal tetrad
\begin{eqnarray*}
\mbox{\boldmath$\theta^{\alpha}$} = \sqrt{\mid g_{\alpha\alpha} \mid}
\mbox{\boldmath$dx^{\alpha}$}, \hspace{2cm} \mbox{(no summation over $\alpha$)}
\end{eqnarray*}
for each metric (all four line-elements are
diagonal).

\subsection{Analysis of the solution A}
After rescaling the coordinates $\{t,x,y,z\}$ and a
coordinate change $e^{t} \cosh(x) \rightarrow t,\;\;$ 
$e^{t} \sinh(x) \rightarrow x$
the metric can be written as 
\[ds^2 = S^{-2}
  \left(- dt^2 + dx^2 + t^{1-a-b}dy^2 + t^{1-a+b}dz^2\right) \]
where $S=t^{-a/2} x^{1-a^2/q}$ and
$a, b$ are essential constants that are free apart from the condition, that
$q$ defined as $q=b^2+a^2-2a-1$ has to be non-zero.
Coordinates $t,x$ are restricted to $0<t$ and $0<x$ to allow arbitrary
real exponents in $S.$
We will see below that in order to satisfy positivity conditions
of energy, the parameter $a$ has to take values for which there is
a curvature singularity at $t=0$ such that the restriction
$t>0$ is no loss of generality.
As a switch $b \leftrightarrow -b$ is equivalent to $y \leftrightarrow z$,
we consider only $b \geq 0.$
The metric has two conformal Killing vectors
\begin{eqnarray*}
\vec{k_1} = \partial_x, \hspace{1cm}
\vec{k_2} = 2t \partial_t + 2x \partial_x + 
               (1+a+b)y \partial_y + (1+a-b)z \partial_z. 
\end{eqnarray*}
With the further abbreviation $D = q^2x^2 - 4a^2t^2$ the fluid velocity 
one-form has components $u_{\alpha}=-\,\mbox{sign}(q)D^{-1/2}(qx, 2at, 0, 0)$. 
As announced before, lower
Greek indices denote components in the frame $\mbox{
\boldmath${\theta}^{\alpha}$}$.
An overall minus sign was chosen in order to have a future pointing $\vec{u}$.
To exclude null or tachyonic fluid the coordinates have to be restricted to
guarantee $D>0$, i.e.
\begin{equation}
 \frac{x}{t}\; > 2\left| \frac{a}{q} \right| . \label{res1}
\end{equation}
The Petrov type of the metric is $I$ and 
the non-vanishing Weyl spinor components
$\Psi$ are
\[\Psi_0 = \Psi_4 = -\, \frac{abS^2}{4t^2},\;\;\; 
  \Psi_2 = \frac{(b^2+a-1)S^2}{12t^2}. \]
In the case $b=0,\,a=1$ the metric is conformally flat.
Then the metric has 6 KVs, shows $\rho - p < 0$ 
and obeys an equation of state which is
\[\rho + p = 
\frac{(p-\rho)}{8}\left[\left(\frac{p-\rho}{12}\right)^2-1\right].\]
It therefore is a special FLRW spacetime.
Otherwise, in any one of the four cases
\[ a=0,\;\;b=0,\;\;b=1,\;\;b=\;\mid 1-a \mid \]
the metric is of Petrov type D.
The expansion of the fluid is 
\[ \theta = \frac{S}{2|q|txD^{3/2}} \,
            [48(a^2-q)a^3t^4-16a(a^2-q)q^2t^2x^2+(a+2)q^4x^4].  \]
The acceleration and the acceleration scalar of the fluid are
\begin{eqnarray*}
 a_{\alpha} & = & -\,\frac{(2a+q)q^2x^2S}{D^2} \left( 2at, qx, 0, 0 \right), \\
 a_{\alpha}a^{\alpha} & = & (2a+q)^2q^4x^4S^2/D^3.
\end{eqnarray*}
The non-vanishing components of the shear tensor $\sigma_{\alpha\beta}$ read
\[ \sigma_{00} = \frac{2at}{qx}\,\sigma_{01} = 
                              \frac{4a^2t^2}{q^2x^2}\,\sigma_{11} = 
   -\,\frac{4a^2|q|txS[D(1-a)-4at^2(2a+q)]}{3D^{5/2}}, \]
\[ \sigma_{22} = 
   \frac{|q|xS[(1-a-3b)D-4(2a+q)at^2]}{6tD^{3/2}},\]
\[ \sigma_{33} = 
   \frac{|q|xS[(1-a+3b)D-4(2a+q)at^2]}{6tD^{3/2}}.\]
The shear scalar is
\[\sigma^{\alpha\beta}\sigma_{\alpha\beta} =
\frac{q^2x^2S^2}{6t^2D^3}\left(3b^2D^2+[4at^2(a^2+b^2-1)+(a-1)D]^2\right).\]
With the energy density 
\[ \rho = \frac{(a^2-q)S^2[q^2x^2-12(a^2-q)t^2]}{4q^2t^2x^2} \]
and the pressure 
\[ p = \frac{(a^2-q)S^2[q^2x^2+4(a^2-3q)t^2]}{4q^2t^2x^2} \]
we obtain
\[ \rho+p = \frac{(a^2-q)S^2D}{2q^2t^2x^2}, \;\;\;\;\;\; 
   \rho-p = \frac{2(a^2-q)(3q-2a^2)S^2}{q^2x^2}. \]
To have the last two quantities non-negative and therefore $\rho$
non-negative, $a$ and $b$ have to satisfy
\[ a \geq -1/2, \;\;\; 2a+1 \geq b^2 \geq 2a+1-a^2/3. \]
For $a>-2$ there is a big-bang like singularity at a finite distance.
There are two possibilities to have an equation of state $p=p(\rho)$:
\begin{itemize}
\item[-] For $q = -2a$ both, $\rho$ and $p$, are functions of $x/t$.
\item[-] For $q=2a^2/3$ we have $p = \rho = (a^2x^2-9t^2)/(12t^{a+2}x^3).$
\end{itemize}

The metric A has an abelian three dimensional conformal subgroup and
therefore belongs to a class of metrics discussed previously by
A.\,M.\,Sintes in \cite{AlTs}
and A.\,M.\,Sintes, J.\,Carot and A.\,A.\,Coley in 
\cite{COA} although no explicit expressions for the metric were
given there. In \cite{COA} it is explained that the
energy conditions $\rho+p>0,\;\rho-p>0$ can only be fulfilled in
a special subclass to which metric A does not belong to.
In contradiction we found that for any $a > -1/2$ there exists 
a range of values
for $b$ such that both energy conditions are satisfied in the
complete perfect-fluid range.

\subsection{Analysis of the Solution B}
The metric is 
\[
ds^2 = \frac{1}{ S^2} \left [
 -\frac{dt^2}{\sinh^{2+2m} (x)}  + \frac{dx^2}{\sinh^{2+2m} (x)}
 + \frac{e^{-2t}}{\sinh^2 (x)} dy^2 +
\mbox{cotanh}^2 (x) dz^2 \right  ], \]
where $S$ depends on the variable $u \equiv
t + \log \sinh (x)$. This coordinate approaches asymptotically a null
coordinate for big $x$.
The conformal Killing vector of this
metric is
\begin{eqnarray*}
\vec{k} = \partial_t + y \partial_y.
\end{eqnarray*}
We define a new function $z(u)$  as
\begin{eqnarray*}
   \frac{dS}{du} \equiv  \left ( z -1 \right) S.  \label{defz}
\end{eqnarray*}
The ordinary differential equation that $S(u)$ had to satisfy has in general
two constants of integration. However, by adding appropriate
constants to the coordinates $t$ and $x$ and rescaling all the coordinates,
it can be easily seen that there only appear two inequivalent cases
(we are discarding the homothetic solutions because they have been already
investigated in \cite{Ali2}).
The explicit expressions for $S$ and $z$ are:
\begin{eqnarray*}
S & = & b e^{-\frac{1+m}{2} u} \cosh (\frac{\sqrt{1+m^2}}{2} u) \\
& & \hspace{2cm} \Rightarrow \hspace{3mm} 
z = \frac{1}{2} \left [ 1 - m + \sqrt{1+m^2} 
 \tanh \left( \frac{\sqrt{1+m^2}}{2}  u \right) \right ], \\
S & = & b e^{-\frac{1+m}{2} u} \sinh (\frac{\sqrt{1+m^2}}{2} u)  \\
& & \hspace{2cm} \Rightarrow \hspace{3mm}
z= \frac{1}{2} \left [ 1 - m + \sqrt{1+m^2} \mbox{cotanh}
 \left( \frac{\sqrt{1+m^2}}{2} u \right) \right ] .               
\end{eqnarray*}
where $b$ is an arbitrary constant.
So, the metric has two essential parameters:
the constant $b$ gives a global scale to the metric and
the constant $m$ 
measures the departure from the maximally symmetric 
anti-de Sitter spacetime (see
below) and therefore is a measure of the inhomogeneity of the spacetime.

The tetrad components of the fluid velocity one-form of this solution are 
\begin{eqnarray*}
u_{\alpha} = \left (-1, 0 ,0 ,0 \right )
\end{eqnarray*}
so, the metric is given in comoving coordinates.
The non-vanishing Weyl spinor components $\Psi$ are
\begin{eqnarray*}
\Psi_0 = \frac{m S^2}{2}   \sinh^{2m+1} (x) e^{x}
, \hspace{1cm}
\Psi_2 = \frac{m S^2}{6} \sinh^{2m} (x), \\
\Psi_4 = -\frac{m S^2}{2}  \sinh^{2m+1} (x) e^{-x},
\hspace{2cm}
\end{eqnarray*}
so that the Petrov type is I except when the constant $m$ vanishes and 
the metric is conformally flat. In this case, the density and pressure
are $ \rho = - 3 b^2/4$ , $p = 3 b^2/4$ and therefore
the spacetime is anti-de Sitter.

The expansion of the fluid is
\begin{eqnarray*}
\theta = S \sinh^{1+m} (x) \left (2-3 z \right).
\end{eqnarray*}  
The shear tensor of the solution is diagonal and its components are
\begin{eqnarray*}
\sigma_{11} = -\frac{1}{2} \sigma_{22} = \sigma_{33} = 
\frac{S}{3}  \sinh^{1+m} (x), \hspace{5mm} \sigma^{\alpha\beta}\sigma_{\alpha
\beta} = \frac{2S^2}{3}   
\sinh^{2+2m} (x).
\end{eqnarray*}
The acceleration of the fluid is 
\begin{eqnarray*}
\mbox{\boldmath$a$} = - \mbox{cotanh} (x) \left (m+z \right) 
\mbox{\boldmath $dx$}.
\end{eqnarray*}
The density and the pressure are 
\begin{eqnarray*}
\rho =  S^2 \sinh^{2m} (x) \left ( m - 3 z^2 \right ), \hspace{1cm}
p =  S^2 \sinh^{2m} (x)  \left ( 2m + 3 z  \right )  z. 
\end{eqnarray*}
  From these expressions it is clear that the perfect fluid does not
satisfy an equation of state $p=p(\rho)$ for any value of $m$.
The solution with $S$ being a hyperbolic sine has always a
region near $u = 0$ where the
energy-density is negative. Thus, we will concentrate on the solution with
$S$ being a hyperbolic cosine. For this solution the energy-density is
positive everywhere provided the constant $m$ satisfies
\begin{eqnarray*}
m > \frac{4}{3}.
\end{eqnarray*}
The range of variation of the coordinates for this solution is
\begin{eqnarray*}
-\infty < t < \infty, \hspace{1cm} 0 < x, \hspace{1cm} -\infty < y < 
\infty, \hspace{1cm} -\infty < z < \infty.
\end{eqnarray*}
When $u  \rightarrow -  \infty$ the function $S$ tends to infinity
in an exponential way. So, the length of the curve $x=\mbox{const.}$,
$y=\mbox{const.}$ and $z= \mbox{const.}$ from any finite value of
$u$ to
$u = -\infty$ is finite. Furthermore, the energy-density diverges
here and therefore we have a big-bang-like singularity at $t= -\infty$.
On the other hand, we have that $u = + \infty$ is located
at an infinite distance and the same happens with $x=0$,
which is also located at
infinity. The expansion of the fluid is positive everywhere and tends
to infinity when we approach the big bang and to zero in the infinite
future. 

Let us now study in which region the pressure is positive. A trivial 
calculation shows that the pressure is positive in either of the
following regions
\[\tanh \left (\frac{
\sqrt{1+m^2}}{2}  u \right) \geq  \frac{m-1}{\sqrt{1+m^2}}
\hspace{5mm} \mbox{or} \hspace{5mm} 
\tanh \left (\frac{\sqrt{1+m^2}}{2}  u \right) 
\leq - \frac{3+m}{3 \sqrt{1+m^2}}. \]
Furthermore, in the range for $m$ we are interested in  ($m > 4/3$),
we have
\begin{eqnarray*}
\frac{m-1}{\sqrt{1+m^2}} < 1 \hspace{1cm} \mbox{and} \hspace{1cm}
- \frac{3+m}{3 \sqrt{1+m^2}}  > -1, 
\end{eqnarray*}
and, therefore, there always exist two values of $u,$ say $u_0$
and $u_1,$ such that the pressure is positive for
\begin{eqnarray*}
-\infty < u \leq u_0 <0 \hspace{1cm} \mbox{and} 
\hspace{1cm} 0 <u_1 \leq 
u < +\infty,
\end{eqnarray*}
i.e.\ it is positive except for a finite interval around
$u=0$.

\subsection{Analysis of the solution C}
After rescaling $\{t,x,y,z \}$ and a coordinate change
$e^{-x} \rightarrow x$
we find that the metric takes the form
\begin{equation}
ds^2= \frac{1}{S^2(x,t)} \left[ -dt^2 + \frac{dx^2}{x^2} + 
x^{b\left( b + 1\right)}
 \cosh ^{1-b}(t)    dy^2
+ x^{b \left(b-1 \right)}  \cosh ^{1+b}(t)  dz^2 \right ],  
\label{metrC}
\end{equation}
where the function $S(x,t)$ is
\[ S(x,t) =  x^{ \frac{1 + b^2}{2} }   
 + s_0 \mid \sinh t \mid^{\frac{1+b^2}{2}} ,  
\hspace{2cm} s_0 \hspace{3mm} 
\mbox{constant.} \]
As a switch $b \leftrightarrow -b$ is equivalent to $y \leftrightarrow z$,
we consider only $b \geq 0.$
It is convenient to define the two functions $S_x(x)$ and $S_t(t)$ as
\begin{eqnarray*}
S_x \equiv x^{ \frac{1+b^2}{2} }, \hspace{1cm}
S_t \equiv s_0 \mid \sinh t \mid^{\frac{1+b^2}{2}} , 
\hspace{2cm} (S= S_x + S_t).
\end{eqnarray*}
The conformal Killing vector of this solution is
\begin{eqnarray*}
\vec{k} = x \partial_x - \frac{1}{2} b \left (1 + b\right)
 y \partial_y + \frac{1}{2} b \left (1 - b \right) z \partial_z.
\end{eqnarray*}
The non-vanishing Weyl spinor components $\Psi$ are
\[
\Psi_0 = \frac{S^2}{4 \cosh t} b \left ( 1 - b^2 \right ) e^{-t},
\hspace{5mm} 
\Psi_2 = \frac{S^2}{12 \cosh^2 t} \left ( 1 - b^2 \right ),
\hspace{5mm}
\Psi_4 = \frac{S^2}{4 \cosh t} b \left ( 1 - b^2 \right ) e^t.
\]
Consequently, the Petrov type of this metric (when $b \neq 1$ and
$b \neq 0$) is I
everywhere except at the
spacelike hypersurface $\cosh t = \mid  b^{-1} \mid $ where it degenerates
to type D. In the particular case  
$b=  1$, the
metric is conformally flat and the density and pressure of the perfect fluid
are
\begin{eqnarray*}
\rho = 3  s_0^2, \hspace{1cm} p = -3 s_0^2,
\end{eqnarray*}
so that the metric is the de Sitter spacetime. When $b=0$ the 
Petrov type is D everywhere.

Let us now analyze the kinematical quantities and the matter content for
the metric (\ref{metrC}).
The tetrad components of the fluid velocity one-form of this solution is
\begin{eqnarray*}
u_{\alpha} = (-1, 0,0,0)
\end{eqnarray*}
and the metric is written in comoving coordinates.

The fluid expansion is
\begin{eqnarray*}
\theta =  \frac{1 }{2 \cosh t \sinh t} \left [
 \frac{}{}
 2 S \sinh^2 t - 3 S_t\left ( 1 +b^2 \right) \cosh^2 t   \right ].
\end{eqnarray*}       
The non-vanishing components of the shear tensor 
and the shear scalar are
\begin{eqnarray*}
\sigma_{11} = -\frac{S}{3}  \tanh t , \hspace{7mm}
\sigma_{22} = \frac{S}{6}  \left (1 - 3b \right)  \tanh t , \\
\sigma_{33} = \frac{S}{6}  \left (1 + 3 b \right)  \tanh  t,
\hspace{7mm}
\sigma_{\alpha \beta} \sigma^{\alpha \beta} = \frac{S^2}{6} \tanh^2 t
\left ( 1 + 3 b^2 \right).
\end{eqnarray*}
The only non-zero tetrad component of the acceleration is
\begin{eqnarray*}
a_1 = -\frac{b^2+1}{2} S_x.
\end{eqnarray*}
The expressions for the energy-density and pressure read
\begin{eqnarray*}
\rho = \frac{
  \left (b^2 -1 \right) S^2 \sinh^2 t+ 3
\left (b^2+1 \right)^2 S_t^2 \cosh^2 t}{4 \sinh^2 t \cosh^2 t},
\hspace{2cm} \\
p =  \frac{
\left (b^2 -1 \right) S^2 \sinh^2 t +
\left (b^2+1 \right) S_t \cosh^2 t \left [
\frac{}{}2 \left (b^2 -1 \right) S_x -
\left (b^2 + 5 \right) S_t \right]}{4 \sinh^2 t \cosh^2 t}.
\end{eqnarray*}
The perfect fluid does
not satisfy an equation of state $p=p(\rho)$ unless $b=1$ (which
corresponds to the de Sitter spacetime, as discussed above) or
when $s_0 =0$. In the latter case the perfect fluid is a stiff fluid 
($p=\rho$) and the
metric coefficients are separable functions in comoving coordinates. Therefore,
this particular case belongs to the class of metrics studied in \cite{Go}
where all the $G_2$ on $S_2$ 
diagonal stiff fluid metrics
with separable metric coefficients were considered. 

In order to find the ranges of variation of the coordinates $\{ t,x,y,z \}$
which define the spacetime, we must distinguish a number of different
cases depending on the signs of $s_0$ and of $(b^2-1)$. The reason 
is that if 
$b^2-1 > 0$ then the spacetime has no curvature singularity at $t=0$
and it can be continued across $t=0$ to negative
values of $t$. On the other hand, when $0 \leq b^2 < 1$ then the
energy-density blows up at $t=0$ and we have a big bang singularity
there. Thus,
the two possible variation ranges for the coordinate $t$ are
\begin{eqnarray*}
-\infty < t < + \infty , \hspace{1cm} & & \mbox{if} \hspace{3mm}  b^2 > 1 \\
0 < t < + \infty \hspace{11mm} & & \mbox{if} \hspace{3mm} 0 \leq b^2 < 1.
\end{eqnarray*}
Regarding the parameter $s_0$, we must impose that
the function $S$ be non-zero (it can be either positive
or negative as only $S^2$ appears in the metric). When $s_0 \geq 0$,
$x$ can take arbitrarily
large values,
but when  $s_0 < 0$ we have that 
two different variation ranges are possible (they correspond respectively
to the region where $S>0$ and  to the region where $S<0$). 
Thus, the different possibilities for the 
variation range for the coordinate $x$ are
\begin{eqnarray*}
0< x < \infty, \hspace{1cm} & & \mbox{if} \hspace{3mm}  s_0 \geq 0 \\
0 < x <  
(-s_0)^{\frac{2}{1+b^2}} \mid \sinh t \mid  \hspace{1cm} & & \mbox{if}
\hspace{3mm} s_0 < 0 \\
 (-s_0)^{\frac{2}{1+b^2}} \mid \sinh t \mid  
< x < +\infty \hspace{1cm} & & \mbox{if} \hspace{3mm} s_0 < 0.
\end{eqnarray*}
These different variation ranges define completely
different spacetimes, they are not different regions of a single spacetime.
Regarding the coordinates $y$ and $z$, their variation range is, in all
cases,
\begin{eqnarray*}
- \infty < y < + \infty, \hspace{1cm} -\infty < z < + \infty.
\end{eqnarray*}
We can now specify which of these different possibilities 
satisfy $\rho >0$, $\rho- p >0$ and $\rho + p >0$
in the whole spacetime. A simple analysis shows that this can be accomplished
only in one case, namely, when
\[b > 1, \hspace{1cm} s_0 < 0,  \]
\[0 < t < +\infty, \hspace{1cm} 0 < x < 
(-s_0)^{\frac{2}{1+b^2}} \mid \sinh t \mid , \]
\[ -\infty  < y < + \infty,
\hspace{1cm} -\infty < z < + \infty \]
(we have to restrict $t$ to $t>0$ because of the variation range of $x$).
There is no curvature singularity anywhere in the spacetime and so
we come to the question of whether it can be extended to a larger spacetime
or not. The vanishing of $S$ at the boundary
\begin{eqnarray*}
x = (-s_0)^{\frac{2}{1+b^2}} \mid \sinh t \mid 
\end{eqnarray*}
suggests that the boundary is located at an infinite distance and therefore 
the spacetime can not be extended across this boundary 
(a formal proof of completeness is more complicated, however).
Regarding the points with $x=0$ the 
Riemann tensor is regular there. The distance to these points along
a curve $t= \mbox{const.}$, $y= \mbox{const.}$ and $z=\mbox{const.}$ is
divergent, which again suggests that they are located at infinity.
This does not prove, however, that the spacetime is complete although it
is an indication for it. Thus, this spacetime has no singularities and 
satisfies the dominant energy conditions \cite{Hell} everywhere.
Unfortunately, the strong energy condition $\rho+3p \geq 0$ is violated
in some regions of the spacetime.

\subsection{Analysis of the solution D}

The metric is 
\begin{eqnarray}
ds^2= \frac{e^{- (1+a)r }}{S^2(r,t)} \left[ -dt^2 + q(r) dr^2
+ e^{-2t}   dy^2
+ \frac{e^{2ar}}{q(r)}  dz^2 \right ], \label{metrD} 
\end{eqnarray}
where $a$ is an arbitrary parameter and $q(r)$ is given by
\begin{eqnarray}
q(r) \equiv \frac{aA e^{2ar}}{1 + A e^{2ar}}, \hspace{15mm}
A \hspace{2mm} \mbox{non-vanishing constant}
\label{defq}
\end{eqnarray}   
(From the positivity of $q$ follows that $A$ must be negative when
$a < 0$). The subcase $a=0$ is included by setting $A=-1$ and performing the
limit $a \rightarrow 0$ which gives $q= 1/(2r)$.  
The range of variation of the coordinates $y$ and $z$ are
\[-\infty < y < \infty, \hspace{1cm}-\infty < z < \infty, \]
while the range of variation for the coordinate $r$ depends on the sign of
$A$
\[A > 0 \hspace{3mm} \Longrightarrow \hspace{3mm}
 -\infty < r < +\infty , \hspace{1cm} A < 0 \hspace{3mm} \Longrightarrow
\hspace{3mm} r > 0.  \]
The function $S$ in the metric
depends on the variable $v \equiv t+r$ and satisfies the 
differential equation
\begin{equation}
\frac{(a+1)}{4} \frac{S_{,v}}{S} + \frac{(3+a)}{4} \frac{{S_{,v}}^2}{S^2} +
\frac{S_{,v}S_{,vv}}{S^2} + \frac{(a+1)}{4} \frac{S_{,vv}}{S} = 0. \label{2a}
\end{equation}
It can be simplified by defining a new function $w(v)$ as
\begin{eqnarray}
\frac{dS}{dv} \equiv \frac{\left (a+1\right) S}{2 \left (w -1 \right)}
\hspace{1cm} a \neq -1, \label{defw}
\end{eqnarray}
(the case $a=-1$ has to be considered separately but we will not study
it here in any detail as it has negative pressure everywhere).
With this equation (\ref{2a}) becomes 
\begin{eqnarray}
\frac{dw}{dv} = \frac{w \left (a + w \right)}{w+1}, \label{eqw}
\end{eqnarray}
which is trivial to integrate (we will not need the explicit solution, though).
Furthermore, $S$ can
be integrated explicitly in terms of $w$ from (\ref{defw}) as
\begin{eqnarray*}
S = b \frac{\mid w -1 \mid \mid w + a \mid^{\frac{1-a}{2a}}}{\mid w
\mid ^{\frac{1+a}{2a}}},
\end{eqnarray*}
where $b$ is an arbitrary positive constant which gives a global scale
to the spacetime.
This expression for $S$ is also valid when $a=0$ by performing the limit 
$a \rightarrow 0$. The result is
\begin{eqnarray*}
S = b \frac{\mid w -1 \mid}{\mid w\mid} e^{\frac{1}{2w}}.
\end{eqnarray*}
The conformal Killing vector of the metric (\ref{metrD}) reads
\begin{eqnarray*}
\vec{k} = \partial_t + y \partial_y.
\end{eqnarray*}    
The Petrov type is D everywhere with the non-vanishing
Weyl spinor components $\Psi$ given by
\begin{eqnarray*}
\Psi_0 = -3 \Psi_2 = \Psi_4 = \frac{ (a -1 )}{4} S^2 e^{(a+1) r}.
\end{eqnarray*}
In the particular case $a=1$ the metric is conformally flat and the
perfect fluid has $\rho = -3b^2/A$ and $p= 3b^2/A$. Thus,
the spacetime in this limiting case is de Sitter ($A<0$), anti-de Sitter
($A>0$) or Minkowski ($ A= \infty$, which gives $q=1$).

Returning to the
general case $a \neq 1$,
the two repeated null principal directions of the Weyl tensor are
\begin{eqnarray*}
\mbox{\boldmath $l$} = \frac{1}{\sqrt{2}} \frac{e^{- \frac{ (a+1) r}{2}}}{
S} \left ( \mbox{\boldmath $dt$} + e^{-t} \mbox{\boldmath $dy$} \right),
\hspace{1cm}
\mbox{\boldmath $k$} = \frac{1}{\sqrt{2}} \frac{e^{- \frac{ (a+1) r}{2}}}{
S} \left ( \mbox{\boldmath $dt$} - e^{-t} \mbox{\boldmath $dy$}\right ). 
\end{eqnarray*}
The tetrad components of the fluid velocity one-form are 
\begin{eqnarray*}
u_{\alpha} =  \frac{1}{\sqrt{q - w^2}} (-\sqrt{q}, 
w, 0, 0)
\end{eqnarray*}
and therefore the fluid velocity vector
does not lie in the two-plane generated by the
two repeated null principal directions at each point of the spacetime. 

The condition for the existence of a timelike $\vec{u}$ is $q > w^2$ (in
the region of the spacetime where $q = w^2$ the energy-momentum tensor
 represents
a null fluid and in the region $0 < q < w^2$ the matter content is a tachyon
fluid, $q<0$ is forbidden in order to preserve signature).

The expansion of the fluid is
\begin{eqnarray*}
\theta = \sqrt{\frac{q-w^2}{q}}
\frac{Se^{\frac{(1+a)r}{2}}}{2(q-w^2)^2 (w-1)} \left [ \frac{}{}
-3a (q +w^2)(q-w^2) + q^2(4 w^2 - 6w -1) - \right . \\
- \left . \frac{}{} 2q w^2(w-1)(w-2) + 3w^4 \right ],
\end{eqnarray*}        
the non-vanishing components of the shear tensor and the shear scalar are
\[\sigma_{00} = \frac{w^2}{q} \sigma_{11}, \hspace{5mm}
\sigma_{01}= - \frac{w}{\sqrt{q}} \sigma_{11}, \hspace{5mm}
\sigma_{33} + \sigma_{22} + \frac{q - w^2}{q} \sigma_{11} = 0, \]
\[\sigma_{11} = \frac{S}{3}e^{\frac{(1+a)r}{2}}  q^2 \sqrt{\frac{q-w^2}{q}}
\frac{(q+w^2)(w+1)}{(q-w^2)^3}, \]
\[\sigma_{22} = \frac{S}{3}e^{\frac{(1+a)r}{2}}  q \sqrt{\frac{q-w^2}{q}}
\frac{(w^2-2q)(w+1)}{(q-w^2)^2}, \]
\[\sigma_{\alpha\beta} \sigma^{\alpha\beta} = 
\frac{2}{3} e^{(a+1)r} S^2 q (w+1)^2 \frac{(q-w^2)^2 +qw^2}{(q-w^2)^3}.\]
The non-zero tetrad components of the acceleration
and the acceleration scalar of the fluid are
\[a_0 = - \frac{w}{\sqrt{q}} a_1 = 
 \frac{ e^{\frac{(1+a)r}{2}} S w^2 }{(q-w^2)^2(w^2-1)}
\left [ \frac{}{} 2 a (q -w^2) - q (w^3+w^2 -3w -1) - 2 w^3 \right ],\]
\[ a_{\alpha} a^{\alpha} = a_1^2 \frac{(q-w^2)}{q}. \] 
The energy-density and the pressure read
\begin{eqnarray*}
\rho\! & = &\! \frac{e^{(1+a)r}S^2}{4(w-1)^2 q}
\left [ \frac{}{} 4 q (w-1) \left ( aw +2w  + 2a +1 \right) +
 3 (a+1)^2 (q-w^2) \right ] , \\
 p\! & = &\! \frac{e^{(1+a)r}S^2}{(w-1)^2 q}
\left [ \frac{}{}  q (1-w) \left ( aw +2w +2a + 1 \right)
-  \frac{ (a+1) (q-w^2) \left ( 5aw + w + a + 5 \right)}{4(w+1)} 
\right ].
\end{eqnarray*}
It can be easily seen that the perfect fluid does not satisfy an
equation of state $p=p(\rho)$ 
unless $a=1$ (i.e.\ the de Sitter, anti-de Sitter and Minkowski
limits of the family, see above).
The expression for $\rho + p$ is simple and reads
\[ \rho + p =  \frac{S^2}{2} e^{(1+a)r} 
   \frac{(q-w^2)(a^2-1)}{(1- w^2)q} \]
and, therefore, in the region where the fluid velocity is timelike we have that
\begin{eqnarray*}
\rho+p >0 \hspace{5mm} \Longleftrightarrow \hspace{5mm}
\frac{a^2-1}{1-w^2} > 0. 
\end{eqnarray*}
In order to study the range of variation of the function $w$ it is convenient
to consider the metric (\ref{metrD}) using the coordinates $\{w,r,y,z \}$. We
will only need the coefficient in $dw^2$ which reads
\begin{eqnarray*}
- e^{-(1+a)r} 
\frac{ (w+1)^2}{b^2 (w-1)^2} \frac{ \mid w \mid ^{\frac{1-a}{a}}}{
\mid w + a \mid ^{\frac{a+1}{a}}} d w^2.
\end{eqnarray*}
Analyzing this coefficient together with the expression for $\rho + p$
we find that the function $w$ is allowed to vary in five different ranges.
The extrema of the intervals are
\[ w \rightarrow -\infty, \hspace{5mm} w= -1, \hspace{5mm} 
w= 0, \hspace{5mm} w= 1, \hspace{5mm} w=a, \hspace{5mm} w \rightarrow 
+ \infty \]
where $a$ may take an arbitrary value.
Each one of the five possibilities gives a different
behaviour for the metric and the matter content. Let us restrict
the possibilities by imposing $\rho + p>0$ and $\rho>0$  
everywhere  in the region
where the matter content is a perfect fluid (that is, in the region $q > w^2$).
It turns out that there exist four different cases in which these two
conditions are fulfilled. The first one is
\begin{eqnarray*}
a < -1 , \hspace{1cm} -1 < w < 0.
\end{eqnarray*}
In this case both hypersurfaces $w = -1$ and $w=0$ are curvature singularities
of the spacetime and they are located at a finite distance.
The second possibility is
\begin{eqnarray*}
a < -1 , \hspace{1cm} 0 < w < 1.
\end{eqnarray*}
Here, $w=0$ is a singularity of the spacetime located at a finite distance
and $w=1$ is at an infinite distance in the future. The third one is
\begin{eqnarray*}
-1 < a < 1, \hspace{1cm} w < -1,
\end{eqnarray*}
now the spacetime has a future singularity at $w=-1$ at a finite distance
and $w= -\infty$ is at the infinite past. Finally,
\begin{eqnarray*}
-1 < a <1, \hspace{1cm} w >1, 
\end{eqnarray*}
which has a curvature singularity at $w=1$ at a finite time in the past
and extends infinitely
into the future.

\section*{Acknowledgements}

The authors wish to thank A. Koutras for the help with the computer algebra
program CLASSI.
M.M wishes to thank J. Senovilla for helpful suggestions related with
this work and the Ministerio de Educaci\'on y Ciencia
for financial support.

\section{Appendix}

The aim of this appendix is the analysis of the equation
\begin{eqnarray}
\Sigma_1 (x) + \h \Sigma_2(x) + \frac{\dot{H}^2}{H^2} \Sigma_3(x) +
\frac{\ddot{H}}{H} \frac{\dot{H}}{H} \Sigma_4(x) + \frac{\ddot{H}}{H}
\Sigma_5(x) =0,\label{q2again}
\end{eqnarray}
which involves two independent variables, $x$ and $t+bK(x)$. Let us
first assume that $\Sigma_4 \equiv 0$. Then the equation reads
\begin{eqnarray*}
\Sigma_1 (x) + \h \Sigma_2(x) + \frac{\dot{H}^2}{H^2} \Sigma_3(x) +
\frac{\ddot{H}}{H} \Sigma_5(x) =0.
\end{eqnarray*}
If, in addition, $\Sigma_5 \equiv 0$ then this equation is a polynomial
of second degree in $\dot{H}/H$ which can be solved (unless the
polynomial is identically zero) giving $\dot{H}/H$ (which
is a function only of $t+b K(x)$) in terms of functions of only $x$. This
implies that $\dot{H}/H$
must be constant, thus producing homothetic solutions, 
against our assumptions. Thus, the only possibility is
\begin{eqnarray*}
\Sigma_1 = \Sigma_2 = \Sigma_3 = \Sigma_4 = \Sigma_5 = 0, \hspace{1cm}
H \hspace{5mm} \mbox{arbitrary}
\end{eqnarray*}
which we label as Case 1.

When $\Sigma_4 \equiv 0$ and $\Sigma_5 \neq 0$, we can divide equation
(\ref{q2again}) by $\Sigma_5$ and apply the differential operator
$\vec{D} \equiv \partial_x - b K' \partial_t$ which kills all functions
depending only on the variable $t+bK(x)$. The result is 
\begin{eqnarray*}
\left ( \frac{\Sigma_1}{\Sigma_5} \right )' + 
\left ( \frac{\Sigma_2}{\Sigma_5} \right )' \h +
\left ( \frac{\Sigma_3}{\Sigma_5} \right )' \frac{\dot{H}^2}{H^2} =0.
\end{eqnarray*}
Now, the same considerations made before imply that each coefficient in
this polynomial in $\dot{H}/H$ must be zero. Thus, we necessarily have
\begin{eqnarray*}
\Sigma_4 =0 , \hspace{5mm} \Sigma_1 = m_1 \Sigma_5, \hspace{5mm}
\Sigma_2= m_2 \Sigma_5, \hspace{5mm} \Sigma_3= m_3 \Sigma_5, \hspace{5mm}
\Sigma_5 \neq 0, \\
m_1 + m_2 \h + m_3 \frac{\dot{H}^2}{H^2} + \frac{\ddot{H}}{H} =0, \hspace{2cm}
\end{eqnarray*}
where $m_1$, $m_2$ and $m_3$ are arbitrary constants. We label this
possibility as Case 2.
We have finished the analysis when $\Sigma_4 \equiv 0$, so let
us assume from now on that $\Sigma_4 \neq 0$. Dividing 
(\ref{q2again}) by $\Sigma_4$ and applying the operator $\vec{D}$ on the
resulting equation, we immediately find
\begin{eqnarray}
\left ( \frac{\Sigma_1}{\Sigma_4} \right )' +
\left ( \frac{\Sigma_2}{\Sigma_4} \right )' \h +
\left ( \frac{\Sigma_3}{\Sigma_4} \right )' \frac{\dot{H}^2}{H^2} 
+ \left ( \frac{\Sigma_5}{\Sigma_4} \right )' \frac{\ddot{H}}{H} =0.
\label{last}
\end{eqnarray}
When $\left ( \Sigma_5/\Sigma_4 \right )' \equiv 0$ we
necessarily have
\begin{eqnarray*}
\Sigma_1 = k_1 \Sigma_4, \hspace{5mm} \Sigma_2= k_2 \Sigma_4, \hspace{5mm}
\Sigma_3 = k_3 \Sigma_4, \hspace{5mm} \Sigma_5 = k_5 \Sigma_4,
\hspace{5mm} \Sigma_4 \neq 0 \\
k_1 + k_2 \h + k_3 \frac{\dot{H}^2}{H^2} + \h \frac{\ddot{H}}{H} +
k_5 \frac{\ddot{H}}{H} = 0, \hspace{2cm}
\end{eqnarray*}
($k_1$, $k_2$, $k_3$ and $k_5$ arbitrary constants), which is the
possibility labeled as Case 3. It only remains the situation when
$\left ( \Sigma_5/\Sigma_4 \right )' \neq 0$. We can divide 
equation (\ref{last}) by 
$\left ( \Sigma_5/\Sigma_4 \right )'$ and apply the differential
operator $\vec{D}$ on the resulting equation. We find a polynomial involving
only $\dot{H}/H$ which, as usual, implies that each coefficient 
must vanish. The result is 
\begin{eqnarray*}
\Sigma_1 = k_1 \Sigma_5 + \beta_1 \Sigma_4, \hspace{1cm}
\Sigma_2 = \alpha_2 \Sigma_5 + \beta_2 \Sigma_4, \hspace{1cm}
\Sigma_3 = \alpha_3 \Sigma_5 + k_3 \Sigma_4, \hspace{1cm}
\end{eqnarray*}
where $k_1$, $k_3$, $\alpha_2$, $\alpha_3$, $\beta_1$ and $\beta_2$
are constants. Equation (\ref{last}) is
\begin{eqnarray*}
k_1 + \alpha_2 \h + \alpha_3 \frac{\dot{H}}{H} + \frac{\ddot{H}}{H} = 0.
\end{eqnarray*}
It still remains to impose the original equation (\ref{q2again}) which,
using all this information, reads
\begin{eqnarray*}
\beta_1 + \left ( \beta_2 - k_1 \right ) \h + \left ( 
k_3 - \alpha_2 \right ) \frac{\dot{H}^2}{H} - \alpha_3
\frac{\dot{H}^3}{H^3} = 0.
\end{eqnarray*}
Each coefficient in this polynomial must vanish and we finally find
\begin{eqnarray*}
\Sigma_1 = k_1 \Sigma_5, \hspace{5mm} \Sigma_2 = k_1 \Sigma_4 + k_3
\Sigma_5, \hspace{5mm} \Sigma_3 = k_3 \Sigma_4, \hspace{5mm}
\Sigma_5 \neq 0, \hspace{5mm} \frac{\Sigma_4}{\Sigma_5} \neq
\mbox{const.} \\
k_1 + k_3 \h + \frac{\ddot{H}}{H} =0, \hspace{4cm}
\end{eqnarray*}
which is the last possibility, labeled as Case 4.

\end{document}